\documentclass[11pt,a4paper,oneside]{extarticle}

\usepackage{amsmath}
\usepackage{amssymb}
\usepackage{mathrsfs}
\usepackage{braket}
\usepackage{bm}
\usepackage{bbm}
\usepackage{multirow}
\usepackage{mathabx}

\usepackage{latexsym}
\def\qed{\hfill $\Box$}
\usepackage{ntheorem}
\theoremstyle{break}
\newtheorem{thm}{Theorem}[section]
\newtheorem{lem}[thm]{Lemma}

\newtheorem{prop}[thm]{Proposition}
\newtheorem*{thm**}{Theorem}
\newtheorem*{lem**}{Lemma}
\newtheorem*{cor**}{Corollary}
\newtheorem*{prop**}{Proposition}
\newtheorem*{propapp}{Proposition \ref{classicalp}}

\theorembodyfont{\normalfont}
\newtheorem{defi}[thm]{Definition}
\newtheorem*{defi**}{Definition}
\newtheorem*{pf}{Proof}

\theoremheaderfont{\itshape}
\newtheorem{example}[thm]{Example}
\newtheorem*{example**}{Example}
\newtheorem{rmk}[thm]{Remark}
\newtheorem*{rmk**}{Remark}

\usepackage{appendix}

\usepackage{setspace}

\usepackage{comment}

\usepackage{color}
\usepackage{xcolor}

\usepackage{here}
\usepackage{authblk}

\makeatletter
\renewcommand{\theequation}
{\arabic{section}.\arabic{equation}}
\@addtoreset{equation}{section}
\makeatother

\usepackage{lineno}
\newcommand*\patchAmsMathEnvironmentForLineno[1]{%
	\expandafter\let\csname old#1\expandafter\endcsname\csname #1\endcsname
	\expandafter\let\csname oldend#1\expandafter\endcsname\csname end#1\endcsname
	\renewenvironment{#1}%
	{\linenomath\csname old#1\endcsname}%
	{\csname oldend#1\endcsname\endlinenomath}}%
\newcommand*\patchBothAmsMathEnvironmentsForLineno[1]{%
	\patchAmsMathEnvironmentForLineno{#1}%
	\patchAmsMathEnvironmentForLineno{#1*}}%
\AtBeginDocument{%
	\patchBothAmsMathEnvironmentsForLineno{equation}%
	\patchBothAmsMathEnvironmentsForLineno{align}%
	\patchBothAmsMathEnvironmentsForLineno{flalign}%
	\patchBothAmsMathEnvironmentsForLineno{alignat}%
	\patchBothAmsMathEnvironmentsForLineno{gather}%
	\patchBothAmsMathEnvironmentsForLineno{multline}%
}

\usepackage{graphicx}

\usepackage{hyperref}
\hypersetup{%
 bookmarksnumbered=true,%
 colorlinks=false,%
 setpagesize=false,%
 pdftitle={},%
 pdfauthor={Ryo Takakura},%
 pdfsubject={},%
 pdfkeywords={}
}
\newcommand{\1}{\mbox{1}\hspace{-0.25em}\mbox{l}}

\newcommand{\HH}{\mathcal{H}}
\newcommand{\LL}{\mathcal{L}}

\newcommand{\ang}[1]{\left\langle{#1}\right\rangle}

\newcommand{\ext}{\mathrm{ext}}

\newcommand{\real}{\mathbb R} 
\newcommand{\complex}{\mathbb C} 

\newcommand{\hi}{\mathcal{H}} 
\newcommand{\hik}{\mathcal{K}} 

\newcommand{\A}{\mathsf{A}}

\makeatletter
\def\mathcomposite{%
	\@ifstar
	{\def\@mathcomposite@option{%
			\baselineskip\z@skip\lineskiplimit-\maxdimen}%
		\@mathcomposite}%
	{\let\@mathcomposite@option\offinterlineskip
		\@mathcomposite}}
\def\@mathcomposite{%
	\@ifnextchar[\@@mathcomposite{\@@mathcomposite[0]}}
\def\@@mathcomposite[#1]#2#3#4{%
	#2{\mathchoice
		{\@mathcomposite@{#1}{#3}{#4}\displaystyle{1}}%
		{\@mathcomposite@{#1}{#3}{#4}\textstyle{1}}%
		{\@mathcomposite@{#1}{#3}{#4}%
			\scriptstyle\defaultscriptratio}%
		{\@mathcomposite@{#1}{#3}{#4}%
			\scriptscriptstyle\defaultscriptscriptratio}}}
\def\@mathcomposite@#1#2#3#4#5{%
	\vcenter{\m@th\@mathcomposite@option
		\dimen@\f@size\p@\dimen@#1\dimen@\dimen@#5\dimen@
		\divide\dimen@ 18
		\edef\@mathcomposite@skipamount{\the\dimen@}%
		\ialign{\hfil$#4##$\hfil\cr
			#2\crcr
			\noalign{\vskip\@mathcomposite@skipamount}%
			#3\crcr}}}
\makeatother

\usepackage{framed}

\usepackage[bf]{caption} 
\usepackage[subrefformat=parens]{subcaption}
\begin{document}  
\title{Programming of channels in generalized probabilistic theories}

\author{Takayuki Miyadera\thanks{
miyadera@nucleng.kyoto-u.ac.jp}\ \ and\ Ryo Takakura\thanks{takakura.ryo.27v@st.kyoto-u.ac.jp}}
\affil{Department of Nuclear Engineering 
	\\Kyoto University
	\\Kyoto daigaku-katsura, Nishikyo-ku, Kyoto, 615-8540, Japan
}
\date{}

\maketitle

\begin{abstract}
For a given target system and apparatus described by quantum theory, the so-called quantum no-programming theorem indicates that a family of states called programs in the apparatus with a fixed unitary operation on the total system programs distinct unitary dynamics to the target system only if the initial programs are orthogonal to each other.
The current study aims at revealing whether a similar behavior can be observed in generalized probabilistic theories (GPTs).
Generalizing the programming scheme to GPTs, we derive a similar theorem to the quantum no-programming theorem.
We furthermore demonstrate that programming of reversible dynamics is related closely to a curious structure named a quasi-classical structure on the state space.
Programming of irreversible dynamics, i.e., channels in GPTs is also investigated.
\end{abstract}

\section{Introduction}
In the field of quantum technology such as quantum computation, implementing unitary dynamics to a target system is one of the most important tasks.
The implementation of various quantum gates is significant in general quantum computation, or, more specifically, the implementation of quantum Fourier transformation is a crucial part of Shor's algorithm \cite{nielsen_chuang_2010}.
In analogy with classical computers, Nielsen and Chuang proposed implementing unitary dynamics by means of ``programmable gate array'' \cite{PhysRevLett.79.321}.
In their scenario, an apparatus was considered besides the target system, and the desired unitary dynamics on the target system were implemented by controlling states of the apparatus called ``programs'' and operating a unitary to the total system.
It could be marvelous if there exist an apparatus and a unitary operator on the total system that realize arbitrary unitary dynamics on the system, but such protocol was proved to be mathematically impossible in \cite{PhysRevLett.79.321}.
In fact, there was proved that if $N$ unitary dynamics on the target system can be programmed, then a perfectly distinguishable set of $N$ states in the apparatus are used as the corresponding programs.
This result is known as a \textit{quantum no-programming theorem}, and has been studied extensively: for instance, its optimal protocol for approximate universal programming was found \cite{PhysRevLett.125.210501}, or a no-programming theorem with respect to measurement processes was also studied \cite{postprocess-programming}

In this paper, we study whether such a relation is general between the possibility of programming unitary (reversible) dynamics and the structure of the apparatus.
We extend the programming scheme from quantum theory to \textit{generalized probabilistic theories (GPTs)} \cite{hardy2001quantum,Barnum_nocloning,PhysRevLett.99.240501,PhysRevA.75.032304,PhysRevA.81.062348,PhysRevA.84.012311,Masanes_Muller_2011,barnum2012teleportation,Lami_PhD,Plavala_2021_GPTs,Takakura_PhD}, which are the most general framework of physics, and  investigate how a family of states in an apparatus should behave as programs when programming reversible dynamics in a target system.
It is then proved similarly to the quantum case that if we can implement a pair of distinct reversible dynamics in the system, then the corresponding programs in the apparatus are perfectly distinguishable.
This observation indicates that the quantum behavior observed in the programming scheme is in fact a more general one.
At the same time, when deriving this observation, we find that a curious structure (named a quasi-classical structure) appears in the target system that quantum theory does not have.
Interestingly, we prove that this structure appears in the apparatus in turn when it can program a fixed number of reversible dynamics on an arbitrary target system.
We also discuss another generalization of the quantum setting on the scenario of programming irreversible dynamics, i.e., channels.

This paper is organized as follows.
In Section \ref{sec:GPTs}, we present a brief review on GPTs.
Not only notions for single systems but also fundamentals for bipartite systems are explained there.
In terms of those descriptions, we generalize the scheme of quantum programming to GPTs in Section \ref{sec:reversible}.
In addition to the generalization of the setting originally introduced by Nielsen and Chuang \cite{PhysRevLett.79.321}, where only reversible dynamics were focused, we also consider programming channles in Section \ref{sec:channel}.
There we also give the concrete observations on how to implement channels if we use states in a family of GPTs called regular polygon theories \cite{1367-2630-13-6-063024}.

\section{Generalized probabilistic theories (GPTs)}
\label{sec:GPTs}
In this section, we present a brief review on GPTs according to \cite{Lami_PhD,Plavala_2021_GPTs,Takakura_PhD}.
\subsection{States, effects, and transformations}
A system is specified by its state space.
A \textit{state space} $\Omega$ is a compact convex set in a finite-dimensional Euclidean space $V$ 
such that $\Omega$ satisfies $V=\mathit{span}(\Omega)$ and $\mathit{aff}(\Omega)$ does not contain the origin $O$ of $V$.\footnote{For a subset $A$ of a vector space, its \textit{convex hull} $\mathit{conv}(A)$, {\itshape affine hull} $\mathit{aff}(A)$, and {\itshape linear span} $\mathit{span}(A)$ are given by $\mathit{conv}(A):=\{\sum_{i=1}^{n}\lambda_{i}a_{i}\mid a_i\in A,\ \lambda_{i}\in[0,1],\ \sum_{i}\lambda_{i}=1,\ \mbox{$n$: finite}\}$, $\mathit{aff}(A):=\{\sum_{i=1}^{n}\lambda_{i}a_{i}\mid a_i\in A,\ \lambda_{i}\in\real,\ \sum_{i}\lambda_{i}=1,\ \mbox{$n$: finite}\}$, and $\mathit{span}(A):=\{\sum_{i=1}^{n}\lambda_{i}a_{i}\mid a_i\in A,\ \lambda_{i}\in\real,\ \mbox{$n$: finite}\}$ respectively.
}
We note that only finite-dimensional cases are treated in this paper.
Elements of $\Omega$ are called \textit{states}, and if a state $\omega\in\Omega$ satisfies $\omega_1=\omega_2=\omega$ whenever $\omega=p\omega_1+(1-p)\omega_2$ with $\omega_1, \omega_2\in \Omega$ and $p\in(0, 1)$, then $\omega$ is called a \textit{pure state}.
We denote by $\Omega^{\ext}$ the set of all pure states in $\Omega$, and we call elements in $\Omega\backslash\Omega^{\ext}$ \textit{mixed states}.
The convexity of a state space originates from the physical intuition that probabilistic mixtures of states are possible: if we can prepare two states $\omega_1$ and $\omega_2$, then we can also prepare a state $p\omega_1+(1-p)\omega_2\ (0<p<1)$ through the probabilistic mixture of $\omega_1$ and $\omega_2$ with respective probabilities $p$ and $1-p$ respectively.
Measurements on the system are described by the notion of effects.
An \textit{effect} $e$ is 
a real-valued linear function on $V$ that satisfies $e(\Omega)
\subseteq [0,1]$, and the set of all effects is denoted by $\mathcal{E}
\subset V^*$.
For a state $\omega\in\Omega$ and effect $e\in\mathcal{E}$, the quantity $e(\omega)$ represents the probability of observing some specific outcome when the system is prepared in $\omega$.
We note that in this article we follow the \textit{no-restriction hypothesis} \cite{PhysRevA.81.062348,PhysRevA.87.052131}, which means that all effects are physically valid.
We often write the expression $e(\omega)$ also as $\ang{e, \omega}$ in the following.
The effect $u\in\mathcal{E}$ satisfying $u(\omega)=1$ for all $\omega \in \Omega$ is called the \textit{unit effect}.
An \textit{observable} $E=\{e_x\}_{x\in X}$ is a family of effects satisfying 
$\sum_{x\in X} e_x = u$. 
In this expression, the index set $X$ represents the set of all outcomes observed in the measurement of $E$, and each effect $e_x$ outputs the probability of observing the outcome $x\in X$ acting on states. 
In this article, we assume that the outcome set of an observable is a finite set.
A family of states $\{\omega_x\}_x\subset\Omega$ is called \textit{perfectly distinguishable} if there exists an observable $\{e_x\}_x$ such that $e_{x}(\omega_y)=\delta_{xy}$. 
On the other hand, a family of states $\{\omega_i\}_{i\in\mathcal{I}}\subset\Omega$ with an index set $\mathcal{I}$ is called \textit{pairwise distinguishable} if any pair $\{\omega_i, \omega_j\}$ of its distinct elements is perfectly distinguishable.
Although the two notions above coincide with each other in quantum and classical theories \cite{Barnum_nocloning,Brunner_2014}, they are in general different notions: a perfectly distinguishable set of states is pairwise distinguishable, but the converse does not necessarily hold in GPTs.
For a finite-dimensional state space $\Omega$, while a perfectly distinguishable set of states is seen easily to be a finite set (bounded by the dimension of the vector space $\mathit{span}(\Omega)$), we can prove that a pairwise distinguishable set of states is also a finite set (see Proposition \ref{prop:appA} in \ref{appA}).

Transformations between systems are described through the notion of channels.
For state spaces $\Omega_1$ and $\Omega_2$ whose underlying vector spaces are $V_1$ and $V_2$ respectively, we define the set $\mathcal{C}(\Omega_1, \Omega_2)$ as $\mathcal{C}(\Omega_1, \Omega_2)=\{\Lambda\colon \Omega_1 \to \Omega_2\mid\mbox{affine}\}=\{\Lambda\colon V_1\to V_2\mid\Lambda(\Omega_1)\subseteq \Omega_2,\ \mbox{linear}\}$, and call its elements \textit{channels}.
When $\Omega_1=\Omega_2=\Omega$, we write $\mathcal{C}(\Omega, \Omega)$ simply as $\mathcal{C}(\Omega)$.
Among channels from a system described by $\Omega$ to itself, reversible dynamics are of particular importance.
A channel $\alpha\in\mathcal{C}(\Omega)$ is called a \textit{reversible dynamics} 
if 
it is bijective, and the set of all reversible 
dynamics on $\Omega$ is written as $GL(\Omega)$.
We remark that not all elements of $GL(\Omega)$ are physically realizable: 
in quantum theory, only reversible dynamics described by unitary operators are allowed \cite{Busch_quantummeasurement,heinosaari_ziman_2011}.

We have so far explained transformations in terms of state changes (the Schr\"odinger picture), but we can also describe them through transitions between effects (the Heisenberg picture).
For a channel $\Lambda\in\mathcal{C}(\Omega_1, \Omega_2)$, its dual map $\Lambda^*\colon \mathcal{E}_2\to\mathcal{E}_1$, where $\mathcal{E}_1$ and $\mathcal{E}_2$ are the respective effect spaces for $\Omega_1$ and $\Omega_2$, is defined by the affine map satisfying
$\langle \Lambda^* (e), \omega\rangle 
= \langle e, \Lambda (\omega)\rangle$ for all $\omega\in\Omega_1$ and $e\in\mathcal{E}_2$.
We note that the dual map $\Lambda^*$ can be extended linearly to $\Lambda^*\colon V_2^*\to V_1^*$, where $V_1^*$ and $V_2^*$ are the dual space of the underlying vector spaces of $\Omega_1$ and $\Omega_2$ respectively, and that the dual $\alpha^*$ of a reversible dynamics $\alpha\in GL(\Omega)$ becomes a bijection on $\mathcal{E}$ (and $V^*$) as well.

\subsection{Bipartite systems}
In this part, the description of bipartite systems in GPTs is briefly reviewed.
Let $\Omega_{1}$ and $\Omega_{2}$ be state spaces embedded in finite-dimensional Euclidean spaces $V_{1}$ and $V_{2}$ respectively,
 and let $\mathcal{E}_{1}$ and $\mathcal{E}_{2}$ be the respective effect spaces for $\Omega_{1}$ and $\Omega_{2}$.
We remember that $\mathcal{E}_{1}$ and $\mathcal{E}_{2}$ are subsets of the dual spaces $V^*_1$ and $V^*_2$ of $V_{1}$ and $V_{2}$ respectively, and $V^*_1\simeq V_1$ and $V^*_2\simeq V_2$ hold due to the assumption of finite dimensionality.
For a bipartite system composed of systems with state spaces $\Omega_{1}$ and $\Omega_{2}$, we write its state space by $\Omega_{12}$.
Then, requiring several physical principles, we obtain the following observations (see \cite{Lami_PhD,Plavala_2021_GPTs,Takakura_PhD} for detailed explanations).
\begin{enumerate}
\item The bipartite state space $\Omega_{12}$ is embedded in the Euclidean space $V_1\otimes V_2$, that is, $\mathit{span}(\Omega_{12})=V_1\otimes V_2$ (thus the bipartite effect space $\mathcal{E}_{12}$ is a subset of $V_1^*\otimes V_2^*$);
\item When states $\omega\in\Omega_1$ and $\xi\in\Omega_2$ are prepared independently in each single system, the bipartite state is given by $\omega\otimes\xi$;
\item When effects $e\in\mathcal{E}_1$ and $f\in\mathcal{E}_2$ are measured independently in each single system, the bipartite effect is given by $e\otimes f$;
\item The unit effect for $\Omega_{12}$ is given by $u_1\otimes u_2$, where $u_1$ and $u_2$ are the unit effects for $\Omega_{1}$ and $\Omega_{2}$ respectively;
\item The bipartite state space $\Omega_{12}$ satisfies
\begin{align}
\label{eq:bipartite}
\Omega_{1}\otimes_{min}\Omega_{2}
\subseteq\Omega_{12}\subseteq\Omega_{1}\otimes_{max}\Omega_{2},
\end{align}
where 
\begin{equation}
\label{eq:min tensor}
\begin{aligned}
\Omega_{1}\otimes_{min}\Omega_{2}
=\{\mu\in V_1\otimes &V_2\mid\mu=\sum_{i=1}^{n}p_i \omega_i\otimes\xi_i,\ \omega_i\in\Omega_{1},
\\&\xi_i\in\Omega_2,\ p_i\ge0,\ \sum_{i=1}^{n}p_i=1,\ n:\mbox{finite}\}
\end{aligned}
\end{equation}
 and
\begin{equation}
\label{eq:max tensor}
\begin{aligned}
\Omega_{1}\otimes_{max}\Omega_{2}
=\{\mu\in V_1\otimes &V_2\mid\ang{u_1\otimes u_2, \mu}=1,
\\&\ \ \langle e\otimes f, \mu\rangle\ge0,\ \mbox{for all $e\in\mathcal{E}_{1}, f\in\mathcal{E}_{2}$}\};
\end{aligned}
\end{equation}
\item The bipartite effect space $\mathcal{E}_{12}$ satisfies
\begin{align}
\mathcal{E}_{1}\otimes_{min}\mathcal{E}_{2}
\subseteq\mathcal{E}_{12}\subseteq\mathcal{E}_{1}\otimes_{max}\mathcal{E}_{2},
\end{align}
where $\mathcal{E}_{1}\otimes_{min}\mathcal{E}_{2}$ and $\mathcal{E}_{1}\otimes_{max}\mathcal{E}_{2}$ are defined in the same way as \eqref{eq:min tensor} and \eqref{eq:max tensor} respectively.
\end{enumerate}
The convex sets $\Omega_{1}\otimes_{min}\Omega_{2}$ and $\Omega_{1}\otimes_{max}\Omega_{2}$ in \eqref{eq:min tensor} and \eqref{eq:max tensor} are called the \textit{minimal tensor product} and the \textit{maximal tensor product} of $\Omega_1$ and $\Omega_2$ respectively (similarly for $\mathcal{E}_{1}$ and $\mathcal{E}_2$).
These convex sets do not coincide with each other unless either state space is classical (a simplex) \cite{Aubrun2021}.
We note that if the bipartite state space $\Omega_{12}$ is given by $\Omega_{12}=\Omega_{1}\otimes_{min}\Omega_{2}$, then the corresponding effect space $\mathcal{E}_{12}$ is $\mathcal{E}_{12}=\mathcal{E}_{1}\otimes_{max}\mathcal{E}_{2}$, and if $\Omega_{12}=\Omega_{1}\otimes_{max}\Omega_{2}$, then $\mathcal{E}_{12}=\mathcal{E}_{1}\otimes_{min}\mathcal{E}_{2}$ holds.
We also remark that these tensor products are compatible with the notion of complete positivity \cite{Plavala_2021_GPTs}.
That is, for a channel $\Lambda\in\mathcal{C}(\Omega_1, \Omega_2)$ and an arbitrary state space $\Omega'$, 
\[
[\Lambda\otimes\mathrm{id_{\Omega'}}](\Omega_{1}\otimes_{min}\Omega')\subseteq\Omega_2\otimes_{min}\Omega'
\]
and
\[
[\Lambda\otimes\mathrm{id_{\Omega'}}](\Omega_{1}\otimes_{max}\Omega')\subseteq\Omega_2\otimes_{max}\Omega'
\]
hold, where $\mathrm{id}_{\Omega'}$ is the identity channel on $\Omega'$ and $\Lambda\otimes\mathrm{id_{\Omega'}}$ is the tensor product of the linear maps $\Lambda$ and $\mathrm{id_{\Omega'}}$.
For a bipartite state space $\Omega_{12}$, we can introduce the notion of \textit{partial trace}.
In fact, we can prove that there exists $\xi\in \Omega_2$ for a bipartite state $\mu\in\Omega_{12}$ such that 
\[
\ang{u_1\otimes f, \mu}_{12}=\ang{f, \xi}_2\quad(\forall f\in\mathcal{E}_2)
\]
holds, where $u_1$ is the unit effect for $\Omega_1$ and $\ang{\cdot, \cdot}_{12}$ and $\ang{\cdot, \cdot}_{2}$ represent the action of an effect on a state in $\Omega_{12}$ and $\Omega_{2}$ respectively (the same observation can be obtained also for $\Omega_1$).
In the following, when considering a bipartite state space composed of state spaces $\Omega_1$ and $\Omega_2$, we often use the tensor product notation $\Omega_1\otimes \Omega_2$ instead of $\Omega_{12}$ to represent the bipartite state space.
We remark that the symbol $\otimes$ used in the expression $\Omega_1\otimes \Omega_2$ does not have any specific meaning as in the tensor product of two vector spaces.

\subsection{Fidelity in GPTs}
How ``close'' two states are can be quantified by means of fidelity in classical and quantum theory \cite{nielsen_chuang_2010}.
In this part, we show that fidelity also can be introduced in GPTs, and present its properties.
The quantity plays a crucial role to prove our main results.

For a state space $\Omega$ and a pair of states $\omega,\sigma \in \Omega$, the fidelity between them is defined by \cite{Zander_2009,KIMURA2010175}
\begin{align}
\label{eq:fidelity}
F(\omega, \sigma):= \inf 
\sum_{x\in X} \langle e_x, \omega\rangle^{1/2}
\langle e_x, \sigma\rangle^{1/2}, 
\end{align} 
where the infimum is taken over all the observables $\{e_x\}_{x\in X}$ on $\Omega$.
The fidelity takes a value in $[0,1]$. 
For states $\omega,\sigma\in\Omega$,
$F(\omega, \sigma)=1$ holds if and only if $\omega=\sigma$. 
On the other hand, $F(\omega, \sigma)=0$ holds if and only if  
there exists an effect $e$ such that $\langle e, \omega\rangle =1$
and $\langle e, \sigma\rangle =0$, i.e., $\{\omega, \sigma\}$ is perfectly distinguishable.
In addition to these observations, the fidelity has the following properties \cite{Zander_2009,KIMURA2010175}.
\begin{prop}
	\label{prop:fidelity}
	Let $\Omega$ and $\Omega'$ be state spaces.\\
(i) $F(\Lambda(\omega), \Lambda(\xi))
\geq F(\omega, \sigma)$ holds for an arbitrary channel $\Lambda\in \mathcal{C}(\Omega, \Omega')$ and states $\omega, \xi\in\Omega$.\\
(ii) $F(\Lambda(\omega), \Lambda(\xi))
	= F(\omega, \xi)$ holds for an arbitrary reversible dynamics $\Lambda\in GL(\Omega)$ and states $\omega, \xi\in\Omega$.\\
(iii)
	$F(\omega_1\otimes \xi_1, 
	\omega_2\otimes \xi_2)
	\leq F(\omega_1, \omega_2) F(\xi_1, \xi_2)$ holds for arbitrary states $\omega_1, \omega_2\in\Omega$ and $\xi_1,\xi_2\in\Omega'$.\\
(iv)
	$F(\omega_1\otimes \xi, \omega_2\otimes \xi)
	= F(\omega_1, \omega_2)$ holds for arbitrary states $\omega_1, \omega_2\in\Omega$ and $\xi\in\Omega'$.
\end{prop}
We note that the expression \eqref{eq:fidelity} of fidelity in GPTs reduces to the usual one when quantum or classical theory is considered \cite{nielsen_chuang_2010,Zander_2009}.

\section{Programming of reversible dynamics in GPTs}
\label{sec:reversible}
In this section, we study how to program reversible dynamics in GPTs generalizing the idea of programming quantum dynamics in \cite{PhysRevLett.79.321}.
\subsection{Programming in quantum theory}
We first review the quantum scenario discussed in \cite{PhysRevLett.79.321}.
Suppose that there exist a quantum system and a quantum apparatus associated with finite-dimensional Hilbert spaces $\hi$ and $\hik$ respectively.
We consider programming a unitary (i.e., reversible) dynamics on the system by choosing a state of the apparatus.
Let $W$ be a unitary operator on $\HH\otimes\hik$.
We say that a state $|\xi\rangle \in \hik$ (called a program) in the apparatus implements a unitary dynamics $U_{\xi}$ on the system through $W$ if the following condition holds for any $|\varphi \rangle \in \hi$: 
\begin{align}
	\label{eq:3}
	W(|\varphi\rangle \otimes |\xi\rangle) 
	= (U_{\xi}|\varphi\rangle) \otimes |\xi'\rangle, 
\end{align} 
where $|\xi'\rangle \in \hik$ is a state of the apparatus.
In \cite{PhysRevLett.79.321}, it was proved that when programs $\xi$ and $\eta$ implement unitary operators $U_{\xi}$ and $U_{\eta}$ respectively, $U_{\xi}\neq U_{\eta}$ is possible only if $\langle \xi | \eta\rangle =0$ holds. 
It implies that the number of programs is at most the dimension of $\hik$. 
We note that one can program $\dim \hik$ number of distinct unitary dynamics $\{U_n\}_{n=1}^{\dim \hik}$ by choosing the unitary operator $W$ to be $W=\sum_n U_n \otimes |n\rangle \langle n |$, where $\{\ket{n}\}_n$ is an orthonormal basis of $\hik$. 
The original proof for the necessity of $\langle \xi | \eta \rangle =0$ 
goes as follows. First it is shown in \eqref{eq:3} that $|\xi'\rangle$ does not 
depend on $|\varphi\rangle$. In fact, assuming 
\begin{align*}
&W|\varphi_1\rangle \otimes |\xi\rangle 
= U_{\xi}|\varphi_1\rangle \otimes |\xi_1'\rangle
\\
&W|\varphi_2\rangle \otimes |\xi\rangle 
= U_{\xi}|\varphi_2\rangle \otimes |\xi_2'\rangle,
\end{align*} 
we take their inner product to obtain  
\begin{align*}
\langle \varphi_1 | \varphi_2\rangle 
= \langle \varphi_1 |\varphi_2 \rangle 
\langle \xi_1' | \xi_2'\rangle. 
\end{align*}
Since the above equality holds for an 
arbitrary pair of nonorthogonal 
$|\varphi_1\rangle$ and $|\varphi_2\rangle$, 
it follows that $\langle \xi_1' | \xi_2'\rangle =1$. 
Now we have 
\begin{align*}
&W|\varphi\rangle \otimes |\xi\rangle 
= U_{\xi}|\varphi\rangle \otimes |\xi'\rangle\\
&W|\varphi\rangle \otimes |\eta\rangle 
=U_{\eta}|\varphi\rangle \otimes |\eta'\rangle. 
\end{align*}
Their inner product indicates 
\begin{align*}
\langle \xi | \eta\rangle= 
\langle \varphi | U_{\xi}^* U_{\eta}|\varphi\rangle
\langle \xi' | 
\eta'\rangle. 
\end{align*}
It implies that the term $\langle \varphi | U_{\xi}^* U_{\eta}|\varphi\rangle$ does not depend on $|\varphi\rangle$ 
unless $\langle \xi | \eta \rangle 
= \langle \xi' | \eta' \rangle =0$. 
Thus for 
nonorthogonal programs $|\xi\rangle$ and $|\eta\rangle$ 
we find $U_{\xi}^* U_{\eta} = c\1$ 
for some $c\in \complex$ and
$|c|=1$ due to the unitarity. 
That is, the program states must be orthogonal with each other.

\subsection{Programming in GPTs}
\label{subsec: program in GPTs}
Let us formulate a similar problem in GPTs.
We have a system and an apparatus associated with state spaces $\Omega_{sys}$ and $\Omega_{app}$ respectively.
The total system is described by their tensor product 
$\Omega_{tot}:= 
\Omega_{sys} \otimes \Omega_{app}$.
We introduce a subset $GL_0(\Omega_{sys})$ of $GL(\Omega_{sys})$ such that any $\alpha \in GL_0(\Omega_{sys})$ satisfies $\alpha \otimes \mbox{id}_{\Omega_{app}} \in GL(\Omega_{sys}\otimes \Omega_{app})$ for the bipartite system $\Omega_{sys}\otimes \Omega_{app}$.
This condition is not satisfied for every $\alpha' \in GL(\Omega_{sys})$ in general (e.g. the transpose map in quantum theory with the usual composite rule), while in the minimal and maximal tensor product
$\alpha' \otimes \mbox{id}_{\Omega_{app}}
\in GL(\Omega_{sys} \otimes_{min} \Omega_{app})$ and 
$\alpha' \otimes \mbox{id}_{\Omega_{app}}
\in GL(\Omega_{sys}\otimes_{max} \Omega_{app})$ holds respectively for any $\alpha'\in GL(\Omega)$.
We also assume that the subset $GL_0(\Omega)$ has a group structure 
with respect to the concatenation, that is, the identity channel $\mathrm{id}_{\Omega_{sys}}\in GL_0(\Omega)$, $\alpha_1\circ \alpha_2 
\in GL_0(\Omega)$ whenever $\alpha_1, \alpha_2\in GL_0(\Omega)$, and $\alpha^{-1}\in GL_0(\Omega)$ whenever $\alpha\in GL_0(\Omega)$. 

Now let us consider a reversible dynamics 
$\Lambda\in GL(\Omega_{tot})$ on the total system.
We say that a state (called a program) 
$\xi \in \Omega_{app}$ implements a
reversible dynamics $\alpha_{\xi}
\in GL_0(\Omega_{sys})$ on $\Omega_{sys}$ through 
$\Lambda$ if the following equation
holds for any $\omega \in \Omega_{sys}$ and $e\in \mathcal{E}_{sys}$: 
\begin{align}
\label{eq8}
\langle e\otimes u_{app}, 
\Lambda(\omega \otimes \xi)\rangle
=\langle e, \alpha_{\xi}\omega\rangle,
\end{align}
where $u_{app}$ is the unit effect for $\Omega_{app}$.
The condition \eqref{eq8} implies that the dynamics restricted on the system coincides with $\alpha_{\xi}$.
In particular, if we consider a pure state $\omega\in 
\Omega_{sys}^{ext}$ of the system,  
then \eqref{eq8} indicates that the state after the reversible evolution is written as 
\begin{align}
\label{eq:programming for pure}
\Lambda(\omega \otimes \xi)
=\alpha_{\xi} \omega \otimes \xi'
\end{align}
with some $\xi'\in \Omega_{app}$ because 
the pure state $\alpha_{\xi}\omega$ cannot have 
any correlation with the apparatus \cite{Namioka_Phelps}.
On the other hand, it follows from a similar observation that programs can be assumed to be pure.
To see this, let $\xi\in\Omega_{app}$ be a program that can be decomposed into a convex combination as $\xi= \sum_n p_n \sigma_n$ with $\sigma_n \in \Omega_{app}^{ext}$. 
Then, from \eqref{eq:programming for pure}, it holds for any $\omega\in \Omega_{sys}^{ext}$ that 
\begin{align*}
	\sum_n p_n 
	\Lambda(\omega\otimes \sigma_n)
	=\alpha_{\xi}\omega \otimes \xi', 
\end{align*}
where $\xi'\in\Omega_{app}$.
Because the restriction (partial trace) of the left hand side to the system is a pure state $\alpha_{\xi}\omega$, we can find that $\Lambda(\omega\otimes \sigma_n)
=\alpha_{\xi}\omega \otimes \sigma'_n$ 
holds with some state $\sigma'_n\in\Omega_{app}^{ext}$.
Thus hereafter we assume programs to be pure.

We investigate conditions that enable 
distinct reversible dynamics to be programmed in GPTs. 
While the original proof for the quantum setting relies on the inner product of the Hilbert spaces (see the last subsection), fidelity introduced in \eqref{eq:fidelity} plays a crucial role to mimic the argument in its generalization to GPTs.
Let $\xi\in \Omega_{app}^{ext}$ be a program 
implementing $\alpha_{\xi}\in GL_0(\Omega_{sys})$ through $\Lambda\in GL(\Omega_{tot})$. 
The following lemma is important.
\begin{lem}
\label{lem:indistinguishable}
Let $\xi\in \Omega_{app}^{ext}$ be a program 
implementing $\alpha_{\xi}\in GL_0(\Omega_{sys})$ through $\Lambda\in GL(\Omega_{tot})$. 
For distinct pure states $\omega_1, \omega_2\in\Omega_{sys}^{ext}$ of the system, define states $\xi_1', \xi_2'\in \Omega_{app}^{ext}$ 
of the apparatus by
\begin{align*}
	&\Lambda(\omega_1\otimes \xi) = \alpha_{\xi}
	\omega_1 \otimes \xi_1',\\
	&\Lambda(\omega_2 \otimes \xi)=\alpha_{\xi}
	\omega_2 \otimes \xi_2'.
\end{align*} 
If $\omega_1$ and $\omega_2$ are not perfectly distinguishable (i.e., the fidelity between them is nonzero), then $\xi_1'=\xi_2'$.
\end{lem}
\begin{pf}
From Proposition \ref{prop:fidelity}, it follows that
\begin{align*}
	F(\omega_1, \omega_2)
	&= F(\omega_1\otimes \xi, 
	\omega_2 \otimes \xi)\\
	&= F(\Lambda(\omega_1 \otimes \xi), \Lambda(\omega_2 \otimes \xi))\\
	&=F(\alpha_{\xi}\omega_1 \otimes \xi_1', 
	\alpha_{\xi}\omega_2\otimes \xi_2')\leq F(\omega_1, \omega_2) F(\xi_1', \xi_2'). 
\end{align*}
Thus we find that $F(\xi_1', \xi_2')=1$, i.e., $\xi_1'=\xi_2'$ holds for $F(\omega_1, \omega_2) \neq 0$.\qed
\end{pf}
Based on this lemma, we can introduce a disjoint 
decomposition $\Omega_{sys}^{ext}= \bigcup_{\nu\in Z}
\Omega_{sys}^{ext}(\nu)$ of $\Omega_{sys}^{ext}$ by the following rule.
We define a binary relation $\sim$ on 
$\Omega^{ext}_{sys}$ by $\omega_1 \sim \omega_2$
if and only if either $F(\omega_1, \omega_2)\neq0$ or 
there exists a set of elements $\{\sigma_1, \ldots, \sigma_{L-1}\}$ of $\Omega^{ext}_{sys}$ such that $F(\sigma_l, \sigma_{l+1}) \neq 0$ holds for every $l=0,1, \cdots, L-1$ (here we set $\sigma_0=\omega_1$ and $\sigma_{L}=\omega_2$).
It is easy to see that this relation is an equivalence relation, and that $\alpha \omega_1
\sim \alpha \omega_2$ if and only if $\omega_1 \sim 
\omega_2$ for $\alpha\in GL(\Omega)$.
Then we obtain a disjoint decomposition $\Omega_{sys}^{ext}=
\bigcup_{\nu \in Z}\Omega_{sys}^{ext}(\nu)$, 
where $Z=\Omega_{sys}^{ext}/\sim$ is the quotient set and $\Omega_{sys}^{ext}(\nu)=\{\omega\mid\omega\in\nu\}$ is the set of all elements of $\Omega_{sys}^{ext}$ that belong to an equivalence class $\nu\in Z$.
It can be shown that when $\nu, \varphi\in Z$ $(\nu \neq \varphi)$,
\begin{align}
\label{eq:same outputs}
\begin{aligned}
&\Lambda(\omega_1\otimes \xi) = \alpha_{\xi}
\omega_1 \otimes \xi',\\
&\Lambda(\omega_2 \otimes \xi)=\alpha_{\xi}\omega_2 \otimes \xi'
\end{aligned} 
\end{align}
hold for any $\omega_1, \omega_2\in \Omega_{sys}^{ext}(\nu)$ 
with $\xi'\in\Omega_{app}^{ext}$ due to Lemma \ref{lem:indistinguishable}, and that $F(\omega_\nu, \omega_\varphi)=0$ holds for any $\omega_\nu \in \Omega_{sys}^{ext}(\nu)$ and
$\omega_\varphi \in \Omega_{sys}^{ext}(\varphi)$ since $\omega_\nu\nsim\omega_\varphi$.
Moreover, the latter observation implies that a family of states $\{\omega_\nu\}_{\nu\in Z}$ with each $\omega_\nu\in\Omega_{sys}^{ext}(\nu)$ is pairwise distinguishable, and thus $Z$ is a finite set.
\begin{example}
	\label{eg:classical}
	A classical system with $N$ pure states 
	is described by a simplical state space with $N$ 
	extreme points. 
	All pure states are inequivalent in this case. 
\end{example}
\begin{example}
	\label{eg:quantum}
	Consider a quantum system described by a 
	Hilbert space $\hi=\complex^d$ ($d<\infty$). 
	Its state space is the set 
	$\Omega=\mathcal{S}(\hi):=\{\rho | \
	\rho \in \LL(\hi), \rho\geq 0, \mbox{tr}[\rho]=1\}$ of all density operators on $\hi$, where $\LL(\hi)$ is the set of all linear operator on $\hi$.
	In this case, all states are equivalent. 
\end{example}
\begin{example}
	\label{eg:super selection}
	Consider a classical-quantum hybrid system 
	whose state space $\Omega$ is described by a direct sum
$\Omega=\bigoplus_{n=1}^N \mathcal{S}(\hi_n)$, where $\mathcal{S}(\hi_n)$ is the quantum state space with a finite-dimensional Hilbert space $\hi_n$ (see Example \ref{eg:quantum}).
The associated observable algebra is given by 
$\mathfrak{A}=\bigoplus_{n=1}^N \LL(\hi_n)$. 
The system is sometimes called a quantum system with a superselection rule. 
In this case, there are $N$ inequivalent classes.
\end{example}
\begin{example}
	\label{eg:square}
	Consider a system described by a square state space. 
	The square has four pure staets $\omega_1, \omega_2,\omega_3, 
	\omega_4$, where $\omega_1$ and $\omega_3$ form a diagonal. 
	In this case all four pure states are inequivalent. 
\end{example}
%

\begin{figure}[h]
	\centering
	\includegraphics[scale=0.33]{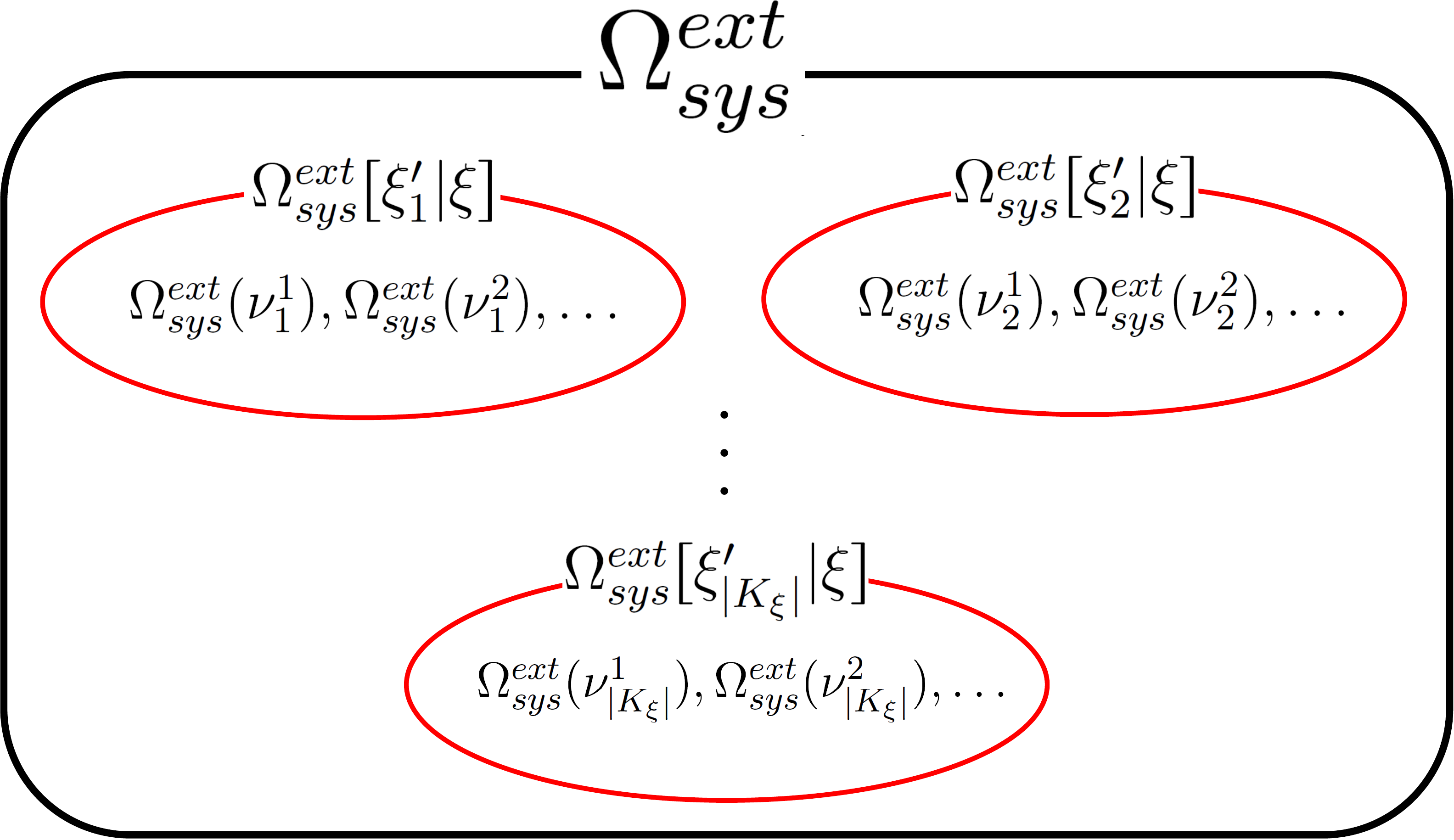}
	\caption{Associated with a program $\xi\in\Omega_{app}$, two decompositions for $\Omega_{sys}^{ext}$ can be introduced.}
	\label{Fig:decompositions}
\end{figure}
Besides the decomposition $\Omega_{sys}^{ext}=
\bigcup_{\nu \in Z}\Omega_{sys}^{ext}(\nu)$, we introduce another decomposition of $\Omega_{sys}^{ext}$ 
based on the program $\xi\in \Omega_{app}^{ext}$. 
Let us define a set $K_{\xi}\subset
\Omega_{app}^{ext}$ by 
\begin{align}
K_{\xi}=\{\xi'\in\Omega_{app}^{ext}\mid\Lambda(\omega
\otimes \xi) 
= \alpha_{\xi} \omega\otimes 
\xi',\ \omega \in \Omega_{sys}^{ext}\}. 
\end{align}
In addition, for each $\xi'\in K_{\xi}$ 
we define $\Omega_{sys}^{ext}[\xi'|\xi]
\subset \Omega_{sys}^{ext}$ by 
\begin{align}
\Omega_{sys}^{ext}[\xi'|\xi]
:= \{\omega\in\Omega_{sys}^{ext}\mid\Lambda(\omega \otimes \xi)
=\alpha_{\xi} \omega \otimes \xi'\}. 
\end{align}
Thus we obtain a disjoint decomposition  
$\Omega_{sys}^{ext}
=\bigcup_{\xi' \in K_{\xi}}\Omega_{sys}^{ext}[\xi'|\xi]$ of $\Omega_{sys}^{ext}$ (see Figure \ref{Fig:decompositions}).
We can find from Lemma \ref{lem:indistinguishable} that if
$\omega_i, \omega_i'\in\Omega_{sys}^{ext}(\nu_i)$ $(\nu_i\in Z)$, i.e., $\omega_i\sim \omega_i'$, then $\omega_i, \omega_i'\in\Omega_{sys}^{ext}[\xi'_i|\xi]$ holds for some $\xi'_{i}\in K_{\xi}$.
It follows that there in general exist $\Omega_{sys}^{ext}(\nu^1_i), \Omega_{sys}^{ext}(\nu^2_i), \ldots$ with $\nu^1_i, \nu^2_i, \ldots \in Z$ for $\xi'_{i}\in K_{\xi}$ such that $\Omega_{sys}^{ext}(\nu^1_i), \Omega_{sys}^{ext}(\nu^2_i), \ldots\subseteq\Omega_{sys}^{ext}[\xi'_i|\xi]$.
That is, the decomposition $\Omega_{sys}^{ext}=\bigcup_{\nu \in Z}\Omega_{sys}^{ext}(\nu)$ is finer than $\bigcup_{\xi' \in K_{\xi}}\Omega_{sys}^{ext}[\xi'|\xi]$ (the latter decomposition will be studied further in the next section).
We note that we can in particular obtain $|K_{\xi}|<\infty$ because $Z$ is finite (see the argument above Example \ref{eg:classical}). 

Now we investigate how programs should be organized to implement dynamics in GPTs.
Let us first consider the case where all elements of $\Omega_{sys}^{ext}$ are equivalent to each other, i.e., $|K_{\xi}|=1$.
In this case, 
if we consider two distinct programs $\xi, \eta \in \Omega_{app}^{ext}$, then 
it holds for any $\omega\in \Omega_{sys}$ that
\begin{align*}
&\Lambda(\omega\otimes \xi) = \alpha_{\xi}\omega \otimes
\xi',\\
&\Lambda(\omega\otimes \eta) = \alpha_{\eta}\omega
\otimes \eta'
\end{align*}
with $\xi', \eta'\in\Omega_{app}$ since $\omega$ can be represented as a mixture of pure states.
Letting $\omega=\omega_0$, where $\omega_0$ is a 
fixed point for $GL(\Omega_{sys})$ (see e.g. \cite{Takakura2020} for its construction), in the above equations, we obtain 
\begin{align*}
&\Lambda(\omega_0\otimes \xi) = \omega_0\otimes \xi',\\
&\Lambda(\omega_0\otimes \eta) = \omega_0\otimes \eta'. 
\end{align*}
Again the properties of fidelity are applied to show 
\begin{align*}
F(\xi, \eta)&=F(\omega_0 \otimes \xi, \omega_0 \otimes \eta)\\
&=F(\Lambda(\omega_0 \otimes \xi), 
\Lambda(\omega_0 \otimes \eta))\\
&=F(\omega_0 \otimes \xi', \omega_0 \otimes \eta')=F(\xi', \eta').
\end{align*}
Furthermore, for arbitrary $\omega \in \Omega_{sys}$, 
we find 
\begin{align*}
F(\xi, \eta)&=F(\omega\otimes \xi, \omega\otimes \eta)\\
&= F(\Lambda(\omega\otimes \xi), 
\Lambda(\omega\otimes \eta))\\
&=F(\alpha_{\xi}\omega\otimes \xi', 
\alpha_{\eta}\omega\otimes \eta')\\
&\leq F(\alpha_{\xi}\omega, 
\alpha_{\eta}\omega)
F(\xi', \eta')
=F(\alpha_{\xi}\omega, 
\alpha_{\eta}\omega)
F(\xi, \eta).
\end{align*}
Thus we conclude that 
$\alpha_{\xi}\omega\neq \alpha_{\eta}\omega$ 
is possible only if $F(\xi, \eta)=0$ is satisfied.
\begin{rmk}
In \cite{Wu2013}, a similar mathematical setting and result to the ones in the above argument were shown under a condition called the ``covariant condition'' similar to our $|K_{\xi}|=1$.
In this paper, as we shall demonstrate below, we treat more general cases without the mathematical assumption of $|K_{\xi}|=1$, and observe that the conclusion $F(\xi, \eta)=0$ holds also in those general cases.
\end{rmk}

Let us next consider the general case where a nontrivial decomposition 
$\Omega_{sys}^{ext}=\bigcup_{\theta'\in K_{\theta}}
\Omega_{sys}^{ext}[\theta'|\theta]$ may exist for some program $\theta$. 
For distinct programs $\xi,\eta\in\Omega_{app}^{ext}$ implementing $\alpha_{\xi}, \alpha_{\eta}\in GL_0(\Omega_{sys})$ through $\Lambda\in GL(\Omega_{tot})$, 
suppose first that there exists a pure state $\omega_{\nu}\in 
\Omega_{sys}^{ext}(\nu)$ of the system
satisfying $\alpha_{\xi}\omega_{\nu} 
\nsim
\alpha_{\eta}\omega_{\nu}$.
Since $\xi,\eta\in\Omega_{app}^{ext}$ are programs, it holds that 
\begin{equation}
	\label{eq9}
\begin{aligned}
	&\Lambda (\omega_{\nu} \otimes \xi)=\alpha_{\xi} \omega_{\nu} \otimes \xi_{\nu},\\
	&\Lambda(\omega_{\nu}\otimes \eta)=\alpha_{\eta}\omega_{\nu}\otimes \eta_{\nu}. 
\end{aligned}
\end{equation}
with $\xi_{\nu}, \eta_{\nu}\in\Omega_{app}$.
Because $F(\alpha_{\xi}\omega_{\nu}, 
\alpha_{\eta}\omega_{\nu})=0$, we obtain 
\begin{align*}
F(\xi, \eta)&=F(\omega_{\nu}\otimes \xi, 
\omega_{\nu}\otimes \eta)\\
&=F(\Lambda(\omega_{\nu}\otimes \xi), 
\Lambda(\omega_{\nu}\otimes \eta))\\
&=F(\alpha_{\xi}\omega_{\nu}\otimes \xi_{\nu}, 
\alpha_{\eta}\omega_{\nu}\otimes \eta_{\nu})
\leq 0. 
\end{align*}
Thus $F(\xi, \eta)=0$, i.e., $\xi$ and $\eta$ must be distinguishable.
On the other hand, suppose that 
$\alpha_{\xi} \omega
\sim \alpha_{\eta} \omega$ holds 
for all $\omega \in \Omega_{sys}^{ext}$, and pick up an arbitrary pure state $\omega_{\nu}\in \Omega_{sys}^{ext}(\nu)$.  
Defining the reversible dynamics $\Gamma_{\xi}:=(\alpha_{\xi}^{-1} \otimes \mbox{id})
\circ \Lambda\in GL(\Omega_{tot})$, 
we can see from \eqref{eq9} that
\begin{align}
	\label{eq:10}
\begin{aligned}
	&\Gamma_{\xi}(\omega_{\nu}\otimes \xi)
	=\omega_{\nu} \otimes \xi_{\nu},\\
	&\Gamma_{\xi}(\omega_{\nu}\otimes \eta)
	=[(\alpha_{\xi}^{-1}\circ \alpha_{\eta})\omega_{\nu}]
	\otimes \eta_{\nu}
\end{aligned} 
\end{align}
hold.
Let $\Omega_{sys}(\nu)$ denote the convex hull of $\Omega_{sys}^{ext}(\nu)$: $\Omega_{sys}(\nu)=\{\omega\in\Omega_{sys}\mid\omega= \sum_n p_n \omega_{\nu, n}, \omega_{\nu, n}\in \Omega_{sys}^{ext}(\nu), p_n \geq 0, \sum_n p_n =1\}$.
We remember that the condition $\alpha_{\xi} \omega_\nu
\sim \alpha_{\eta} \omega_\nu$ implies $\omega_\nu
\sim (\alpha_{\xi}^{-1}\circ\alpha_{\eta}) \omega_\nu$, and thus $\alpha_{\xi}^{-1} \circ \alpha_{\eta}$ is a
bijection on $\Omega_{sys}^{ext}(\nu)$, which induces an affine bijection on $\Omega_{sys}(\nu)$.
It follows that similar relations to \eqref{eq:10} hold for an invariant state $\omega_{\nu}^{inv}\in \Omega_{sys}(\nu)$ of $GL(\Omega_{sys}(\nu))$: 
\begin{align*}
&\Gamma_{\xi}(\omega_{\nu}^{inv}\otimes \xi)
=\omega_{\nu}^{inv}\otimes \xi_{\nu},\\
&\Gamma_{\xi}(\omega_{\nu}^{inv}\otimes \eta)
= \omega_{\nu}^{inv}\otimes \eta_{\nu}. 
\end{align*}
Comparing the fidelity, we find 
\begin{align}
F(\xi, \eta)=F(\xi_{\nu}, \eta_{\nu}). 
\end{align}
Therefore, for arbitrary $\omega_{\nu}
\in \Omega_{sys}^{ext}(\nu)$, we obtain 
\begin{align*}
F(\xi, \eta)&=F(\omega_{\nu}\otimes \xi, 
\omega_{\nu}\otimes \eta)\\
&= F(\alpha_{\xi}\omega_{\nu}\otimes \xi_{\nu}, 
\alpha_{\eta}\omega_{\nu}\otimes 
\eta_{\nu})\\
&\leq F(\alpha_{\xi}\omega_{\nu}, 
\alpha_{\eta}\omega_{\nu})
F(\xi_\nu, \eta_\nu)=F(\alpha_{\xi}\omega_{\nu}, 
\alpha_{\eta}\omega_{\nu})
F(\xi, \eta).  
\end{align*}
It concludes that $F(\alpha_{\xi}\omega_{\nu}, 
\alpha_{\eta}\omega_{\nu})\neq 1$ is 
possible only if $F(\xi, \eta)=0$. 
We have proved the following theorem. 
\begin{thm}
	\label{thm: program in GPTs}
	If states $\xi$ and $\eta$ of the apparatus 
	implement distinct reversible 
	dynamics of the system, 
	then they are distinguishable. 
\end{thm}
Theorem \ref{thm: program in GPTs} can be considered as a generalization of the quantum result in \cite{PhysRevLett.79.321} to GPTs:
programs in the apparatus should be pairwise distinguishable, and thus only finite number of reversible dynamics can be programmed on the system.

\subsection{Quasi-classical structure}
In this part, we study the decomposition $\Omega^{ext}_{sys}=\bigcup_{\xi' \in K_{\xi}}\Omega^{ext}_{sys}[\xi'|\xi]$ of the pure states of the system introduced by a program $\xi\in\Omega_{sys}^{ext}$ in the previous subsection.
To do this, we need some terminologies.
\begin{defi}
	\label{def:quasi-classical}
	Let $\Omega$ and $\Omega^{ext}$ be the state space of a system and the set of all its pure states respectively, and consider a disjoint decomposition $\Omega^{ext}
	=\bigcup_{z\in \mathcal{Z}}\Omega^{ext}[z]$ of $\Omega^{ext}$ with each $\Omega^{ext}[z]\neq\emptyset$ and $|\mathcal{Z}|\geq 2$. 
	We call the decomposition a
	\textit{quasi-classical decomposition of degree $|\mathcal{Z}|$}
	if there exists 
	an observable $A=\{A_z\}_{z\in\mathcal{Z}}$ called a \textit{quasi-classical observable} satisfying 
	$\langle A_z, \omega_{z'}\rangle = \delta_{zz'}$ for 
	each $\omega_{z'}\in \Omega^{ext}[z]$. 
	A system that yields a quasi-classical decomposition 
	is called to have a \textit{quasi-classical structure}.
\end{defi}
We can prove that the decomposition $\Omega^{ext}_{sys}=\bigcup_{\xi' \in K_{\xi}}\Omega^{ext}_{sys}[\xi'|\xi]$ introduced previously is quasi-classical.
\begin{prop}
	\label{prop:classical substructure}
	Assume that a program $\xi\in \Omega_{app}^{ext}$ 
	implements $\alpha_{\xi}\in GL_0(\Omega_{sys})$ through 
	$\Lambda \in GL(\Omega_{sys}\otimes \Omega_{app})$.  
	If $|K_{\xi}|\ge2$, then the decomposition 
	$\Omega_{sys}^{ext}=\bigcup_{\xi' \in K_{\xi}}
	\Omega_{sys}^{ext}[\xi'|\xi]$ is quasi-classical.  
\end{prop}
\begin{pf}
For each $\xi' \in K_{\xi}$, define $\Omega_{sys}[\xi'|\xi]$ as the convex hull of $\Omega_{sys}^{ext}[\xi'|\xi]$. 
We observe 
$\Omega_{sys}[\xi'|\xi] \cap \Omega_{sys}[\xi''|\xi] 
= \emptyset$ for $\xi' \neq \xi''$ 
since $\omega \in \Omega_{sys}[\xi'|\xi]
\cap \Omega_{sys}[\xi''|\xi]$ implies 
$\Lambda(\omega \otimes \xi)=
\alpha_{\xi}\omega \otimes \xi'
= \alpha_{\xi}\omega \otimes \xi''$. 
Let us consider a reversible dynamics $\Gamma_{\xi}:= (\alpha_{\xi}^{-1}
\otimes \mbox{id})\circ \Lambda: 
\Omega_{tot} \to \Omega_{tot}$, 
which gives for $\omega_{\xi'} \in\Omega_{sys}[\xi'|\xi]$
\begin{align*}
\Gamma_{\xi} (\omega_{\xi'}\otimes \xi)
= \omega_{\xi'} \otimes \xi'. 
\end{align*}
We remember that the channel $\alpha_{\xi}^{-1}
\otimes \mbox{id}$ is assumed to be an element of $GL(\Omega_{tot})$.
For a general $\omega\in \Omega_{sys}$, because it can be decomposed as 
$\omega=\sum_{\xi'\in K_{\xi}} p_{\xi'}
\omega_{\xi'}$ with $\omega_{\xi'}\in \Omega_{sys}[\xi'|\xi]$
and a probability distribution $\{p_{\xi'}\}_{\xi'\in K_{\xi}}$, 
it follows that
\begin{align*}
	\Gamma_{\xi}(\omega\otimes \xi)
	=\sum_{\xi'} p_{\xi'}\omega_{\xi'}
	\otimes \xi'. 
\end{align*}
Thus we can define successfully an affine map 
$\hat{\Gamma}_{\xi}\colon \Omega_{sys} \to \Omega_{sys}
\otimes_{min} \Omega_{app}$ as 
\begin{align*}
	\hat{\Gamma}_{\xi} (\omega) := 
	\Gamma_{\xi}(\omega\otimes \xi). 
\end{align*}
Now we iterate this map. 
We introduce an affine map $\hat{\Gamma}_{\xi}
\otimes \mbox{id}: 
\Omega_{sys}\otimes_{min} 
\Omega_{app} \to (\Omega_{sys}\otimes_{min}
\Omega_{app}) \otimes_{min} \Omega_{app}
=\Omega_{sys}\otimes_{min}\Omega_{app}
\otimes_{min}\Omega_{app}
$, where the right-hand side consists of 
the mixtures of $\omega \otimes \omega_1 \otimes \omega_2$ 
with $\omega\in \Omega_{sys}$ and $\omega_1, \omega_2
\in \Omega_{app}$ \cite{Plavala_2021_GPTs}. 
We note that this affine map is well-defined. 
It can be seen that the map
\begin{align*}
	(\hat{\Gamma}_{\xi}\otimes \mbox{id})\circ \hat{\Gamma}_{\xi}
	: \Omega_{sys} \to \Omega_{sys}\otimes_{min}\Omega_{app}
	\otimes_{min}\Omega_{app}, 
\end{align*}
gives for $\omega_{\xi'}\in \Omega_{sys}[\xi'|\xi]$
\begin{align*}
	\omega_{\xi'} \mapsto 
	\omega_{\xi'}\otimes \xi' \otimes \xi'. 
\end{align*}
We iterate this procedure to obtain 
a map $\Omega_{sys}
\to \Omega_{sys}\otimes_{min}
\Omega_{app}^{\otimes_{min}^M}$ such that
\begin{align*}
	\omega_{\xi'} 
	\mapsto \omega_{\xi'}
	\otimes \xi' \otimes\xi'\otimes\xi' \otimes \cdots 
	\otimes \xi'
\end{align*}
holds for $\omega_{\xi'}\in \Omega_{sys}[\xi'|\xi]$.
Let us consider an observable
$F=\{f_{\xi'}\}_{\xi'\in K_{\xi}}$ on $\Omega_{app}$ (note that $|K_{\xi}|<\infty$) , and write $p(\xi''| \xi')= \langle f_{\xi''}, \xi'\rangle$. 
Due to the assumption of distinctness of $\{\xi'\}_{\xi'\in K_{\xi}}$, the observable $F$ can be chosen so that it distinguishes $\{\xi'\}_{\xi'\in K_{\xi}}$, i.e., the observed probability distributions satisfy $p(\cdot | \xi') \neq p(\cdot |\xi'')$ for $\xi'\neq \xi''$ (see Proposition \ref{prop:appB} in \ref{appB}).
 If we measure an observable 
$\{f_{\xi'_1}\otimes f_{\xi'_2}\otimes 
\cdots \otimes f_{\xi'_M}\}_{(\xi'_1, \cdots, \xi'_M)\in (K_{\xi'})^M}$ on the $M$ apparatuses $\Omega_{app}^{\otimes_{min}^M}$, 
we obtain 
\begin{align}
p(\xi'_1, \xi'_2, \cdots, \xi'_M|\xi')
:= p(\xi'_1|\xi')\cdots p(\xi'_M|\xi'). 
\end{align}
as a probability to observe $(\xi'_1, \cdots, \xi'_M)$.
According to the law of large numbers (Theorem 12.2.1 in \cite{Cover:2006:EIT:1146355}), 
the freqeuncy distribution 
$p_M(\xi'' | \xi'):= |\{n\mid\xi'_n=\xi''\}|/M$ behaves as 
\begin{align}
\mathrm{Prob}\left\{ 
D(p_M(\cdot | \xi') \Vert 
p(\cdot | \xi'))>\varepsilon)
\right\} \leq 2^{-M 
	\left( \varepsilon - |K_{\xi}| \frac{\log(M+1)}{M}\right)}, 
\end{align}
where $D(p\Vert q)$ denotes the relative entropy of probability distributions $p$ and $q$. 
It follows that for sufficiently large $M$ 
the frequency distribution becomes very close to 
$p(\cdot |\xi')$ in almost probability one. 
Thus, by counting the frequency, we can estimate $\{\xi'\}$ with arbitrarily high accuracy. 
Hence, taking $M\to \infty$, we conclude that 
there exists an observable 
$A_{\xi}:= \{A^{\xi}_{\xi'}\}_{\xi'\in K_{\xi}}$ 
satisfying 
$\langle A^{\xi}_{\xi'}, \omega_{\xi''}\rangle
=\delta_{\xi' \xi''}$ for $\omega_{\xi''}\in 
\Omega_{sys}[\xi''|\xi]$ (see Theorem 1 in \cite{Barnum_nocloning} for the mathematically rigorous construction of such $A_{\xi}$). 
\qed
\end{pf}
\begin{example}
	A classical system has a quasi-classical structure. 
\end{example}
\begin{example}
	A quantum system $\mathcal{S}(\complex^d)$ (see Example \ref{eg:quantum}) does not have a quasi-classical structure.
	To see this, suppose that  $\mathcal{S}^{ext}(\complex^d)=\bigcup_{i=1}^K \mathcal{S}_i$ is a quasi-classical decomposition with a quasi-classical observable $A=\{A_i\}_{i=1}^{K}$ $(K\le d)$, where $\mathcal{S}^{ext}(\complex^d)$ is the set of all pure states of $\mathcal{S}(\complex^d)$.
	Then an effect $A_k$ of $A$ should output $0$ or $1$ when acting on an arbitrary pure state.
	However, for pure states obtained by superpositions of elements in $\mathcal{S}_i$ and $\mathcal{S}_j$ $(i\neq j)$, the effect $A_k$ in general does not output $0$ or $1$, which is a contradiction.	
\end{example}
\begin{example}
	A quantum system with a superselection rule (see Example \ref{eg:super selection}) described by a state space $\Omega=\bigoplus_{n=1}^N \mathcal{S}(\hi_n)$  has a quasi-classical structure: $\Omega^{ext}=\bigcup_{n=1}^N \mathcal{S}^{ext}[n]$ with
	\begin{align*}
		&\mathcal{S}^{ext}[1]=\mathcal{S}^{ext}(\hi_1)\oplus0\oplus0\oplus\cdots,\\
		&\mathcal{S}^{ext}[2]=0\oplus\mathcal{S}^{ext}(\hi_2)\oplus0\oplus0\oplus\cdots,\\
		&\qquad\qquad\vdots\\
		&\mathcal{S}^{ext}[n]=0\oplus0\oplus\cdots\oplus0\oplus\mathcal{S}^{ext}(\hi_n).
	\end{align*}
	In this case, with $\1_n$ the identity operator on $\hi_n$, the observable $\{\1_n\}_n$ gives a quasi-classical observable.	
\end{example}
\begin{example}
	For a family of state spaces $\Omega[n]\subset V_n$ ($n=1,2,\ldots,N$), define $\Omega:=\bigoplus_n \Omega[n]\subset \bigoplus_n V_n$ by a direct sum: $\Omega=\{\oplus_n p_n \omega_n|\ \omega_n \in \Omega[n], 
	p_n \geq 0, \sum_n p_n =1\}$. 
	Then the state space $\Omega$ has a quasi-classical structure in a similar way to the previous example.
\end{example}
\begin{example}
	A square system in Example \ref{eg:square} has a quasi-classical structure. 
	It has two distinct decompositions: $\Omega^{ext}=\{\omega_1, \omega_2\} 
	\cup \{\omega_3, \omega_4\}=\{\omega_1, \omega_4\}\cup 
	\{\omega_2, \omega_3\}$.
\end{example}
\begin{example}
\label{eg:triangular prism}
Consider a state space $\Omega$ described by a triangular prism in Figure \ref{Fig:triangular}.
For the set of its pure states $\Omega^{ext}=\{\omega_1, \ldots, \omega_6\}$, we have quasi-classical decompositions 
\begin{align*}
\Omega^{ext}&=\{\omega_1, \omega_2, \omega_3\}\cup\{\omega_4, \omega_5, \omega_6\}\\
&=\{\omega_1, \omega_4\}\cup\{\omega_2, \omega_5\}\cup\{\omega_3, \omega_6\}.
\end{align*} 
\begin{figure}[h]
	\centering
	\includegraphics[scale=0.33]{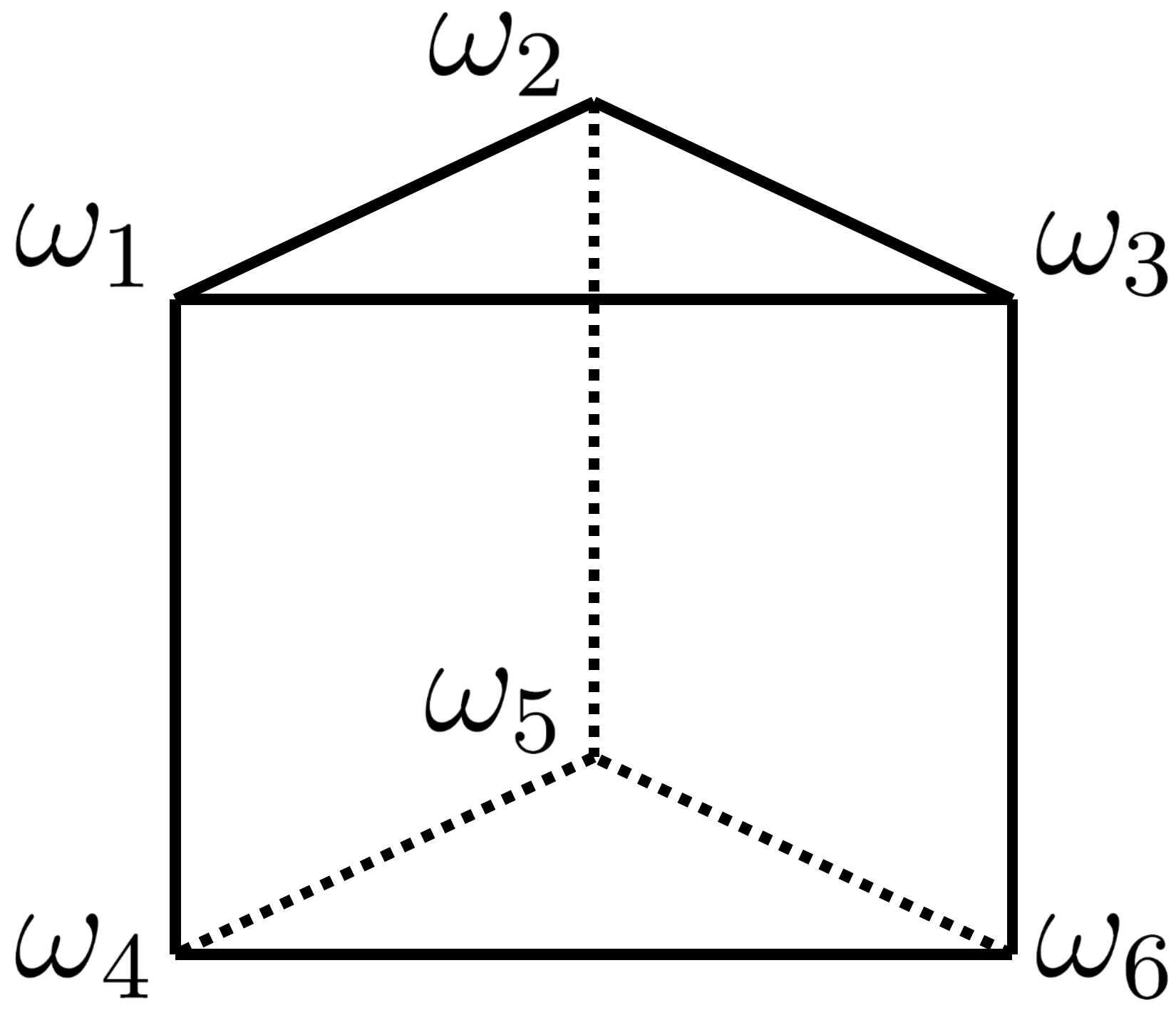}
	\caption{A state space shaped by a triangular prism.}
	\label{Fig:triangular}
\end{figure}
\end{example}
The last two examples show a difference 
between a quasi-classical structure and a classical system. 
For example, in the square system, each state $\omega$ is 
decomposed as $\omega=p \omega_{12} +(1-p)\omega_{34}$, 
where $\omega_{12}$ is 
a mixture of $\omega_1$ and $\omega_2$ and $\omega_{34}$ of $\omega_3$ 
and $\omega_4$. 
For a given $\omega$, while $p$ is uniquely determined, 
$\omega_{12}$ and $\omega_{34}$ are not unique.

To conclude this subsection, we exhibit several properties of quasi-classical structures.
\begin{prop}
	\label{prop:finite quasi}
Let $\Omega$ be a state space with $\dim\mathit{span}(\Omega)=d\ (d<\infty)$ and $\Omega^{ext}$ be the set of all its pure states, 
If there exists a quasi-classical decomposition $\Omega^{ext}=\bigcup_{z\in \mathcal{Z}}\Omega^{ext}[z]$ for $\Omega^{ext}$, then $|\mathcal{Z}|\le d$ holds, and the equality is satisfied only if $\Omega$ is a simplex with $d$ pure states.
\end{prop}
\begin{pf}
To prove the first claim, we introduce a set of its elements $\{\omega_z\}_{z\in \mathcal{Z}}$ with each $\omega_z\in\Omega^{ext}[z]$ and consider a relation $\sum_{z\in\mathcal{Z}} c_z\omega_z=0$ $(c_z\in\real)$.
Applying the corresponding quasi-classical observable $A=\{A_z\}_{z\in\mathcal{Z}}$, we find that $c_z=0$ for all $z\in\mathcal{Z}$, i.e, $\{\omega_z\}_{z\in \mathcal{Z}}$ is a linearly independent set, which implies $|\mathcal{Z}|\le d$.
To prove the second claim, assume that $|\mathcal{Z}|=d$ holds.
In this case, we can find that each $\Omega^{ext}[z]$ is composed of only one element.
In fact, if we suppose $\omega_z, \omega'_z\in\Omega^{ext}[z]$, then $\omega'_z$ can be expressed as $\omega'_z=\sum_{z\in Z} c_z\omega_z$ with each $\omega_z\in\Omega^{ext}[z]$ and $c_z\in \real$ (note that such $\{\omega_z\}_{z\in Z}$ is a basis of $V$ due to the assumption $|\mathcal{Z}|=d$).
Then, applying the corresponding quasi-classical observable $A=\{A_z\}_{z\in\mathcal{Z}}$, we find $\omega'_z=\omega_z$.
Hence the pure states of $\Omega$ is given by $d$ linearly independent states.
\qed
\end{pf}
\begin{prop}
	\label{classicalp}
	Let $\Omega$ and $\Omega^{ext}$ be a state space and the set of all pure states respectively, and let $\Omega^{ext}=\bigcup_{z\in \mathcal{Z}}
	\Omega^{ext}[z]$ be a disjoint decomposition for $\Omega^{ext}$. 
	The decomposition $\Omega^{ext}=\bigcup_{z\in \mathcal{Z}}
	\Omega^{ext}[z]$ is quasi-classical if and only if it satisfies the following condition ($\star$):\\
	($\star$)
	if $\omega \in \Omega$ is expressed as $\omega= \sum_{z\in \mathcal{Z}}
	p_z \omega_z=\sum_{z\in \mathcal{Z}}
	q_z \omega'_z$, where each $\omega_z, \omega'_z \in 
	\Omega[z]$ with $\Omega[z]$ the convex hull of $\Omega^{ext}[z]$ and $\{p_z\}_{z\in\mathcal{Z}}$ and $\{q_z\}_{z\in\mathcal{Z}}$ probability 
	distributions on $\mathcal{Z}$, then $p_z=q_z$ holds for all $z\in Z$.
\end{prop}
The proof of this proposition is given in \ref{appC}.
\subsection{Universal programmings in GPTs}
In the previous part, we found that a quasi-classical structure naturally appears in a system when we consider implementing reversible dynamics via a program in an apparatus.
In the following, we show that an apparatus with 
a quasi-classical structure also plays an important role.
\begin{defi}
	\label{def:univ orogram}
	Let 
	$N$ be an integer with $N>0$. 
	An apparatus $\Omega_{app}$ has an \textit{$N$-universal programming property} if 
	for any $\Omega_{sys}$ with $|GL(\Omega_{sys})|\geq N$ 
	and arbitrary $\{\alpha_n\}_{n=1}^N \subset GL(\Omega_{sys})$ 
	there exist 
	a composite system $\Omega_{sys}\otimes \Omega_{app}$ and $\Lambda 
	\in GL(\Omega_{sys}\otimes \Omega_{app})$ such that there are states $\{\xi_n\}_{n=1}^N
	\subset \Omega_{app}^{ext}$ implementing the 
	$N$ reversible dynamics $\{\alpha_n\}_{n=1}^N$ through $\Lambda$.  
\end{defi}
We obtain the following theorem. 
\begin{thm}
	\label{thm:univ program}
	An apparatus $\Omega_{app}$ has the 
	$N$-universal programming property 
	if and only if $\Omega_{app}$ has a quasi-classical structure
	such that $\Omega_{app}^{ext}=
	\bigcup_{z\in \mathcal{Z}}\Omega_{app}^{ext}[z]$ with 
	$|\mathcal{Z}|\geq N$.  
\end{thm}
The following two lemmas prove the claim. 
\begin{lem}
	Let $\Omega_{app}$ be an apparatus with a quasi-classical strucutre $\Omega_{app}^{ext}=
	\bigcup_{n=1}^N \Omega_{app}^{ext}[n]$.	
	For a system $\Omega_{sys}$ with $|GL(\Omega_{sys})|\geq N$ and $\{\alpha_n\}_{n=1}^N \subset 
	GL(\Omega_{sys})$, one can construct 
	a composite system $\Omega_{tot}=\Omega_{sys}\otimes \Omega_{app}$ and 
	a reversible dynamics $\Lambda\in GL(\Omega_{tot})$ such that each pure state $\xi_n \in \Omega_{app}^{ext}[n]$ works as a program implementing its corresponding 
	$\alpha_{n}$. 
\end{lem}
\begin{pf}
Let $A=\{A_n\}_{n=1}^N$ be a quasi-classical observable 
corresponding to the quasi-classical decomposition 
$\Omega_{app}^{ext}
=\bigcup_{n} \Omega_{app}^{ext}[n]$. 
We employ the minimum tensor product to define $\Omega_{tot}:=\Omega_{sys}\otimes_{min}\Omega_{app}$.  
We first note that the composite system $\Omega_{tot}$ also has 
a quasi-classical structure. 
In fact, we can see that the set $\Omega_{tot}^{ext}$ of all its pure states is given by $\Omega_{tot}^{ext}
=\{\omega\otimes \sigma\mid\omega \in \Omega_{sys}^{ext}, \sigma \in \Omega_{app}^{ext}\}$. 
It has a decomposition 
\begin{align*}
	\Omega_{tot}^{ext}
	=\bigcup_n \{\omega\otimes \sigma\mid
	\omega\in \Omega_{sys}^{ext}, 
	\sigma \in \Omega_{app}^{ext}[n]\} 
	=:\bigcup_n \Omega_{tot}^{ext}[n]. 
\end{align*}
This decomposition is quasi-classical because
the observable $u \otimes A:= 
\{u\otimes A_n\}_n$ satisfies 
$\langle u\otimes A_n, \omega \otimes \sigma_{m}\rangle 
=\delta_{nm}$ for $\sigma_m \in \Omega_{app}^{ext}[m]$. 
It follows that each 
$\Theta \in \Omega_{sys}\otimes_{min}\Omega_{app}$ 
is decomposed as 
$\Theta = \sum_n p_n \Theta_n$ with $\Theta_n 
\in \Omega_{tot}[n]$ ($\Omega_{tot}[n]:=\mathit{conv}(\Omega_{tot}^{ext}[n])$). 
The probability distribution 
$\{p_n\}_n$ is uniquely determined due to Proposition \ref{classicalp}.  
Now we define a map 
$\Lambda\colon\Omega_{tot}^{ext}[n]\to\Omega_{tot}^{ext}[n]$ by 
$\Lambda(\omega \otimes \sigma)
=\alpha_n \omega \otimes \sigma$. 
This map can be extended safely to an affine bijection $\Lambda\colon\Omega_{tot}[n]\to\Omega_{tot}[n]$.
We can further extend this map to the whole $\Omega_{tot}$ by  
$\Lambda(\Theta) = \sum_n p_n \Lambda(\Theta_n)$.
It is easy to see that $\Lambda\in GL(\Omega_{tot})$ holds.
\qed
\end{pf}
We write a simplex with 
$M$ pure states as $\Delta_M$: $\Delta_M=\mathit{conv}(\{\delta_m\}_{m=1}^M)$ with affinely independent $\{\delta_m\}_{m=1}^M$. 
\begin{lem}
	\label{lem:univ program2}
	Let $\Omega_{sys}$ be a classical system 
	with $N$ pure states, i.e., $\Omega_{sys}=\Delta_N$, and $\Omega_{app}$ be an apparatus.  
	If for arbitrary $N$ reversible dynamics $\{\alpha_n\}_{n=1}^N$ on $\Omega_{sys}$ there exists a reversible dynamics $\Lambda$ on the composite $\Omega_{sys}\otimes\Omega_{app}$ (remember that $\Omega_{sys}\otimes\Omega_{app}=\Omega_{sys}\otimes_{min}\Omega_{app}$ holds because $\Omega_{sys}$ is a simplex)
	through which $\{\alpha_n\}_{n=1}^N$ are implemented, 
	then $\Omega_{app}$ has a quasi-classical structure with degree $N$. 
\end{lem}
\begin{pf}
Let $\{\delta_m\}_{m=1}^N$ be the pure states of 
$\Omega_{sys}$. 
Each reversible dynamics on $\Omega_{sys}$ is 
described by a permutation of $\{1,\ldots, N\}$.
Let us consider a reversible dynamics $\Lambda$ 
on $\Omega_{tot}=\Omega_{sys}\otimes \Omega_{app}$.
For any $m=1, \ldots, N$ and $\omega\in \Omega_{app}$, 
a pure state $\delta_m \otimes \omega$ 
is mapped to $\Lambda(\delta_m \otimes \omega)
\in \Omega_{tot}^{ext}$.
It is expressed as
$\Lambda(\delta_m \otimes \omega)=\sum_k 
\delta_k\otimes \Lambda_k^m(\omega)$, 
where $\{\Lambda_k^m\}_{k}$ 
are affine maps defined on $\Omega_{app}$.  
Since $\sum_k \delta_k\otimes 
\Lambda_k^m(\omega)$ is pure, there exists $k_0$ (depending on $\omega$) 
such that $\Lambda_{k_0}^m(\omega) \in \Omega_{app}^{ext}$ 
and $\Lambda_{k}^m(\omega)=0$ for $k\neq k_0$. 
Operating $u_{sys}\otimes u_{app}$, we find that
$\sum_{k=1}^N \Lambda^{m*}_k(u_{app}) =u_{app}$ holds, where $\Lambda^{m*}_k$ is the dual map for $\Lambda^{m}_k$ (the Heisenberg picture), that is, $A^m:=\{A^{m}_k\}_{k=1}^N$ with $A^{m}_k=\Lambda^{m*}_k(u_{app})$ is an observable satisfying either $\langle A^m_{k}, \omega\rangle =0$ or $1$ for $\omega \in \Omega_{app}^{ext}$. 
On the other hand, there exists a family of permutations $\{\pi_n\}_{n=1}^N$ 
such that $\pi_n(1)=n$ for each $n=1,\ldots, N$, and we introduce $N$ reversible dynamics $\{\alpha_n\}_{n=1}^N$ 
by $\alpha_n(\delta_m)= \delta_{\pi_n(m)}$ for all $l=1, \ldots, N$. 
Assume that 
$\Lambda\in GL(\Omega_{sys}\otimes\Omega_{app})$ implements $\{\alpha_n\}_{n=1}^N$.
By the assumption, there exist programs $\{\xi_n\}_{n=1}^N\subset\Omega_{app}^{ext}$ such that 
$\Lambda^1_k(\xi_n) =0$ holds except for $k=\pi_n(1)=n$, that is, $\langle A^1_k, \xi_n\rangle = \delta_{k n}$ holds.
Therefore, we conclude that $\Omega_{app}$ has a quasi-classical 
structure of degree $N$. 
\qed
\end{pf}
	We should remember that the degree satisfies $|\mathcal{Z}|\le \dim\mathit{aff}(\Omega_{app})+1$ and that the equality is attained by a classical system with $(\dim\mathit{aff}(\Omega_{app})+1)$ pure states (see Proposition \ref{prop:finite quasi}).


\color{black}
\section{Programming of channels in GPTs}
\label{sec:channel}
We have so far considered programming reversible dynamics in the framework of GPTs to prove that this is possible only when the apparatus is close to classical theory.
In this section, we investigate whether similar observation can be obtained when programming more general state changes, i.e., channels.
\subsection{Irreversible universal programming}
We start with introducing a similar notion to the one in Definition \ref{def:univ orogram}.
\begin{defi}
	Let $N$ be an integer with $N>0$. 
	An apparatus $\Omega_{app}$ has an \textit{irreversible $N$-universal programming property} if
	for any $\Omega_{sys}$ with $|\mathcal{C}(\Omega_{sys})|\geq N$ 
	and arbitrary $\{\tau_n\}_{n=1}^N \subset \mathcal{C}(\Omega_{sys})$,
	there exist 
	a composite system $\Omega_{sys}\otimes \Omega_{app}$ and $\Theta 
	\in \mathcal{C}(\Omega_{sys}\otimes \Omega_{app})$ such that 
	there are states $\{\xi_n\}_{n=1}^N
	\subset \Omega_{app}^{ext}$ implementing 
	$N$ distinct channels $\{\tau_n\}_{n=1}^N$ 
	of $\Omega_{sys}$ through $\Theta$.  
\end{defi}
Similarly to Theorem \ref{thm:univ program}, we obtain the following observation.
\begin{thm}
	\label{thm:irr univ program}
	An apparatus $\Omega_{app}$ has an irreversible
 $N$-universal programming property if and only if 
	there exists a family of perfectly distinguishable states $\{\xi_n\}_{n=1}^{N}$ in $\Omega_{app}$.
\end{thm}
\begin{rmk}
	A similar result to Theorem \ref{thm:irr univ program} was obtained also in \cite{2014209}.
	They both manifest that if we use a set of states as programs to implement arbitrary channels on an arbitrary system via a channel on a total system, then it is necessary and sufficient that the states are perfectly distinguishable.
    This indication can be compared with our previous result Theorem \ref{thm:univ program}, where quasi-classical structures
appear as a consequence of considering reversible dynamics instead of channels.
\end{rmk}
The following two lemmas prove the claim of the theorem.
\begin{lem}
	Let $\{\xi_n\}_{n=1}^{N}\subset\Omega_{app}$ be a perfectly distinguishable set of states, and $\Omega_{sys}$ be a system with $|\mathcal{C}(\Omega_{sys})|\geq N$ 
	and $\{\tau_n\}_{n=1}^N \subset \mathcal{C}(\Omega_{sys})$. 
	One can construct a composite system $\Omega_{tot}=\Omega_{sys}\otimes \Omega_{app}$ and 
	a channel $\Theta\in \mathcal{C}(\Omega_{tot})$ 
	such that each state $\xi_n$ 
	works as a program implementing its corresponding channel $\tau_{n}$. 
\end{lem}
\begin{pf}
Let $A=\{A_n\}_{n=1}^{N}$ be an observable for the perfectly distinguishable $\{\xi_n\}_{n=1}^{N}$ such that $\ang{A_n, \xi_m}=\delta_{mn}$. 
We employ the minimum tensor to define $\Omega_{tot}:=
\Omega_{sys}\otimes_{min}\Omega_{app}$.  
Now we constitute 
$\Theta\in\mathcal{C}(\Omega_{tot})$ 
in the following way.
First, we define a map $\theta_1\colon\Omega_{tot}\to\Omega_{tot}\otimes_{min}\Delta_N$ by the relation $\theta_1(\omega \otimes 
\sigma)=\omega \otimes 
\sigma\otimes\delta$ $(\omega\in\Omega_{sys}, \sigma\in\Omega_{app})$ with some $\delta\in\Delta_N$ and its affine extension, where $\Delta_N$ is a simplex with $N$ pure states $\{\delta_n\}_{n=1}^N$.
We note that this map $\theta_1$ defines successfully a channel: $\theta_1\in\mathcal{C}(\Omega_{tot},\Omega_{tot}\otimes_{min}\Delta_N)$.
Next, we define another map $\theta_2\colon\Omega_{tot}\otimes_{min}\Delta_N\to\Omega_{tot}\otimes_{min}\Delta_N$ as a measure-and-prepare channel \cite{Plavala_2021_GPTs} on $\Omega_{app}\otimes_{min}\Delta_N$ by the observable $\{A_n\otimes u_{\Delta}\}_{n=1}^{N}$ and states $\{\xi'_n\otimes\delta_n\}_{n=1}^{N}$ with $u_{\Delta}$ the unit effect for $\Delta_N$ and $\{\xi'_n\}\subset\Omega_{app}$ states in $\Omega_{app}$.
That is, $\theta_2$ is given by the tensor product of the identity channel on $\Omega_{sys}$ and the corresponding measure-and-prepare channel on $\Omega_{app}\otimes_{min}\Delta_N$, which particuarly satisfies
\[
\theta_2(\omega \otimes 
\sigma\otimes\delta)=\sum_{n=1}^{N}\langle
A_n, \sigma\rangle\ \omega \otimes\xi'_{n}\otimes\delta_n.
\]
Finally, we introduce $\theta_3$ by the tensor product of the identity channel on $\Omega_{app}$ and the channel $\gamma$ on $\Omega_{sys}\otimes_{min}\Delta_N$ defined via 
$\gamma(\omega\otimes\delta_{n})=\tau_{n}(\omega)\otimes\delta_n$ and its affine extension such that $\theta_3(\omega \otimes\xi'_{n}\otimes\delta_n)=\tau_{n}(\omega)\otimes\xi'_n\otimes\delta_n$.
We note 
that each $\theta_i$ $(i=1, 2, 3)$ is not necessarily bijective.
It is easy to see that the map $\Theta\in\mathcal{C}(\Omega_{tot})$ given by the composite of the channel $\theta_3\circ\theta_2\circ\theta_1\in\mathcal{C}(\Omega_{tot},\Omega_{tot} \otimes \Delta_N)$ and the partial trace implements $\{\tau_n\}_n$. 
\qed
\end{pf}
\begin{lem}
	\label{lem:irr univ program2}
	Let $\Omega_{sys}$ be a classical system 
	with $N$ pure states, i.e., $\Omega_{sys}=\Delta_N$.  
	Assume that an apparatus $\Omega_{app}$  
	implements arbitrary 
	$N$ distinct channels on $\Omega_{sys}$ with programs $\{\xi_n\}_{n=1}^{N}\subset\Omega_{app}$.
	Then $\{\xi_n\}_{n=1}^{N}$ is perfectly distinguishable.
\end{lem}
\begin{pf}
Let $\{\delta_m\}_{m=1}^N$ be the pure states of 
$\Omega_{sys}$.
We introduce $N$ channels $\{\tau_n\}_{n=1}^N$ 
by $\tau_n(\delta_m)= \delta_{\pi_n(m)}$ $(m=1, \ldots, N)$, where $\{\pi_n\}_{n=1}^N$ is a family of permutations satisfying $\pi_n(1)=n$ for each $n=1,\ldots, N$.
Assume that states $\{\xi_n\}_n$ programs $\{\tau_n\}_n$ through a channel $\Theta\in \mathcal{C}(\Delta_N\otimes\Omega_{app})$.
We obtain $\Theta(\delta_m\otimes\xi_n)=\tau_{n}\delta_m\otimes\xi'_{mn}=\delta_{\pi_{n}(m)}\otimes\xi'_{mn}$ with $\xi'_{mn}\in\Omega_{app}$ for each pure state $\delta_m\in\Delta_N$ of the system.
The term $\Theta(\delta_m\otimes\xi_n)$ has another expression $\Theta(\delta_m\otimes\xi_n)=\sum_{k=1}^{N}\delta_k\otimes\Theta_{k}^m(\xi_n)
$
through maps $\{\Theta_{k}^m\}_{k,m}$ on $\Omega_{app}$.
It follows that
\begin{align*}
	\Theta_{k}^m(\xi_n)=
	\left\{
	\begin{aligned}
		&\xi'_{mn}\ &&(k=\pi_{n}(m))\\
		&0\ &&(\mbox{otherwise}),
	\end{aligned}
	\right.
\end{align*}
or
\begin{align*}
	\langle \Theta_{k}^{m*}(u_{app}), \xi_n\rangle
	=\left\{
	\begin{aligned}
		&1\quad(k=\pi_{n}(m))\\
		&0\quad(\mbox{otherwise}),
	\end{aligned}
	\right.
\end{align*}
where $u_{app}$ is the unit effect on $\Omega_{app}$ and $\Theta_{k}^{m*}$ is the dual map for $\Theta_{k}^{m}$.
We note that $\sum_{k=1}^N\Theta_{k}^{m*}(u_{app})=u_{app}$ holds, i.e., $\{\Theta_{k}^{m*}(u_{app})\}_{k=1}^N$ defines an observable on $\Omega_{app}$ for each $m$.
Therefore, if we define an observable $\{A_{k}\}_{k=1}^{N}$ with $A_{k}=\Theta_{k}^{1*}$, then it satisfies $\langle A_k, \xi_{n}\rangle=\delta_{kn}$, i.e., $\{\xi_{n}\}_n$ is perfectly distinguishable. 
\qed
\end{pf}

\subsection{Approximate programmings via regular polygon theories}
In the previous subsection, we demonstrated that channels can be programmed by means of perfectly distinguishable set of states in any GPT.
Then it is natural to ask how well one can program them if we use a set of states that are not perfectly distinguishable.
In this part, following the argument in Lemma \ref{lem:univ program2} and Lemma \ref{lem:irr univ program2}, we consider programming $M$ channels $\{\tau_{i}\}_{i=1}^{M}$ on a classical system by using $M$ pure states of the \textit{regular polygon theory} \cite{1367-2630-13-6-063024} with $M$ sides as programs, and investigate how well those programs can realize the desired channels.

The state space $\Omega^{poly}_{M}$ of the regular polygon theory with $M$ sides is given by the convex hull of $M$ pure states $\{\omega^{M}_{i}\}_{i=0}^{M-1}$ in $V=\real^3$ with
\begin{align}
	\label{eq:poly states}
\omega^{M}_{i}=\left(
\begin{array}{c}
r_{M}^2\cos\frac{2\pi i}{M}\\
r_{M}^2\sin\frac{2\pi i}{M}\\
1
\end{array}
\right),
\end{align}
where $r_{M}=[\cos(\frac{\pi}{M})]^{-\frac{1}{2}}$.
The corresponding effect space $\mathcal{E}^{poly}_{M}$ is given by 
\begin{align}
	\label{eq:polygon effect1}
\mathcal{E}^{poly}_{M}=\mathit{conv}(\{e^{M}_{i}\}_{i=1}^{M})
\ \mbox{with}\ 
e^{M}_{i}=\frac{1}{2}\left(
\begin{array}{c}
\cos\frac{(2i-1)\pi}{M}\\
\sin\frac{(2i-1)\pi}{M}\\
1
\end{array}
\right)
\quad\mbox{($M$: even)},
\end{align}
or
\begin{align}
	\label{eq:polygon effect2}
\begin{aligned}
	&\mathcal{E}^{poly}_{M}=\mathit{conv}(\{e^{M}_{i}\}_{i=1}^{M},\  \{u-e^{M}_{i}\}_{i=1}^{M})\\
	&\qquad\qquad\qquad\mbox{with}\ 
	e^{M}_{i}=\frac{1}{1+r_{M}^{2}}\left(
	\begin{array}{c}
		\cos\frac{2\pi i}{M}\\
		\sin\frac{2\pi i}{M}\\
		1
	\end{array}
	\right)
	\quad\mbox{($M$: odd)}.
\end{aligned}
\end{align}
In this expression, we identify effects with elements in the vector space $V=\real^3$ spanned by the state space through the Riesz representation theorem \cite{Conway_functionalanalysis} (thus the action of an effect on a state is given by their inner product).
We also note that under this parameterization, $\mathit{cone}(\mathcal{E}^{poly}_{M})\subseteq\mathit{cone}(\Omega^{poly}_{M})$ holds, i.e., for any $e\in\mathcal{E}^{poly}_{M}\backslash\{0\}$, there exists $\lambda>0$ such that $\lambda e\in\Omega^{poly}_{M}$.

Let us consider the following game 
between two parties, Alice and Bob.
Alice has a classical system $\Delta_N$ with sufficiently large $N$, and hopes to implement a family of reversible dynamics $\{\tau_i\}_{i=1}^M$ on $\Delta_N$.
Representing each $\tau_i$ $(i=1, \ldots, M)$ as $\tau_i(\delta_n) = \delta_{\pi_i(n)}$ through a permutation $\pi_i$ of $\{1,2,\ldots, N\}$, we assume that they satisfy $\pi_i(n) \neq \pi_j(n)$ for all $i\neq j$ and $n\in \{1,2,\ldots, N\}$.
On the other hand, Bob has a polygon 
system $\Omega^{poly}_M$, and can control the total system, that is, Bob can determine channel $\Theta \in \mathcal{C}(\Omega_{tot})$ 
on the total system $\Omega_{tot}=\Delta_N \otimes_{min} \Omega^{poly}_M$.
The game goes as follows: 
Alice chooses randomly one of the dynamics $\tau_i$, and then Bob prepares the corresponding initial state (program) $\omega_i^M$ of $\Omega^{poly}_M$ (see \eqref{eq:poly states}). 
Our question is how well Bob can choose the channel $\Theta$ to make the state $\omega_i^M$ a good program. 
To tackle this problem, let us consider the case where Alice's initial state is $\delta_n\in\Delta_N$.
Because 
\begin{align}
	\label{eq:classical expression0}
	\Theta(\delta_n \otimes \xi) 
	= \sum_{k=1}^N \delta_k \otimes 
	\Theta_k^n(\xi)
\end{align}
holds for all $\xi\in\Omega_M^{poly}$, where $\{\Theta_k^n\}_{k=1}^N$ are affine maps on $\Omega^{poly}_M$, the probability of observing the successful dynamics $\tau_i(\delta_n)=\delta_{\pi_i(n)}$ for this $\delta_n$ is given by $\langle \Theta_{\pi_i(n)}^{n*}(u_{app}), \omega_i^M\rangle$ with $\Theta_k^{n*}$ the dual map for $\Theta_k^{n}$ $(k=1, \ldots, N)$.
The average with respect to the initial state is $\frac{1}{N} \sum_{n=1}^N  \langle \Theta_{\pi_i(n)}^{n*}
(u_{app}), \omega_i^M \rangle$. 
Since Alice chooses the dynamics randomly, 
the total success probability is written as 
\begin{align}
	\label{eq:max prob0}
	P_{suc}(\Theta):= \frac{1}{MN}
	\sum_{i=1}^M \sum_{n=1}^N \langle \Theta_{\pi_i(n)}^{n*}
	(u_{app}), \omega_i^M\rangle. 
\end{align}
Then what we want to obtain is the maximum success probability with respect to every channel on the total system implementing by Bob, that is, 
\begin{align}
	\label{eq:max prob}
	P_M:= \max_{\Theta\in\mathcal{C}(\Omega_{tot})} P_{suc}(\Theta). 
\end{align}

To evaluate \eqref{eq:max prob0}, we focus on the expression \eqref{eq:classical expression0}.
We can find that the maps $\{\Theta_k^n\}_{k=1}^N$ satisfy $\sum_{k=1}^N
\langle u_{app}, \Theta_k^n(\xi)\rangle =1$ 
for all $\xi\in\Omega^{poly}_M$. 
Thus $\Theta$ defines a family of 
$N$ observables $\{\Theta^{n*}(u_{app})\}_{n=1}^N$, where each observable $\Theta^{n*}(u_{app})$ is given by $\Theta^{n*}(u_{app})=\{\Theta_k^{n*}(u_{app})\}_{k=1}^N$ by means of the dual maps $\{\Theta_k^{n*}\}_k$ for $\{\Theta_k^{n}\}_k$. 
On the other hand, 
for a given family of $N$ observables 
$\{A^n\}_{n=1}^N$ with $A^n=\{A_k^n\}_{k=1}^N$, 
one can construct $\Theta\in\mathcal{C}(\Omega_{tot})$ 
such that $A^n=\Theta^{n*}(u_{app})$ for all $n=1, \ldots, N$. 
In fact, if we set $\Theta_k^n(\xi)
:= \langle A_k^n, \xi\rangle\ \xi^n_k$ 
with an arbitrary state $\xi^n_k\in\Omega^{poly}_M$ for each $n,k=1, \ldots, N$, then the map $\Theta$ defined through the expression \eqref{eq:classical expression0} is a channel on $\Omega_{tot}$ and satisfies the condition $A^n=\Theta^{n*}(u_{app})$ for all $n=1, \ldots, N$.
It follows that the maximization over all channels $\mathcal{C}(\Omega_{tot})$ in \eqref{eq:max prob} can be replaced by the maximization over all families of $N$ observables $\{A^n\}_{n=1}^N$ on $\Omega_M^{poly}$ with $N$ outcomes. 
Now \eqref{eq:max prob} becomes
\begin{align*}
	P_M= \max_{\{A^n\}_{n=1}^N\in [\A(N)]^N}
	\frac{1}{MN}
	\sum_{i=1}^M \sum_{n=1}^N
	\langle A^n_{\pi_i(n)}, \omega_i^M\rangle,
\end{align*}
where $\A(N)$ is the set of all observables on $\Omega_M^{poly}$ with $N$ outcomes and $[\A(N)]^N$ is its $N$ products.
Because $\pi_i(n)\neq \pi_j(n)$ $(i\neq j)$ is assumed, the above equation can be rewritten as
\begin{align}
	P_M&= 
	\frac{1}{N} 
	\sum_{n=1}^N 
	\max_{\{A^n\}_{n=1}^N\in [\A(N)]^N} 
	\frac{1}{M} \sum_{i=1}^M
	\langle A^n_{\pi_i(n)}, \omega_i^M\rangle\notag\\
	&= 
	\max_{A\in \A(N)}
	\frac{1}{M}
	\sum_{i=1}^M 
	\langle A_{i}, \omega_i^M\rangle\notag\\
		&=	\max_{A\in \A(M)}
	\frac{1}{M}
	\sum_{i=1}^M 
	\langle A_{i}, \omega_i^M\rangle.
	\label{eq:max prob1}
\end{align}
We note that the last equation is obtained by identifying an observable $A=\{A_1, \ldots, A_n\}$ with $N$ outcomes with observable $\{A_1, \ldots, A_{M-1}, A_M+A_{M+1}+\cdots+A_N\}$ with $M$ outcomes.

The problem is to search how well we can optimize \eqref{eq:max prob1}, that is, to find an observable $A=\{A_{i}\}_{i=1}^M$ that maximizes \eqref{eq:max prob1}.
This problem is equivalent to the so-called state discrimination problem, and there have been studies on this problem not only in quantum theory \cite{Bae_2015} but also in GPTs \cite{PhysRevA.79.062306,e18020039,PhysRevX.9.031053,PhysRevLett.125.150402}.
Here we refer to the results in \cite{PhysRevA.79.062306} by Kimura et al.
According to their results, if we can find a set of states $\{t_{i}\}_{i=1}^M\subseteq\Omega^{poly}_{M}$ and positive numbers $\{\tilde{p}_{i}\}_{i=1}^M$ with each $\tilde{p}_i\in[0,1]$ (called a ``weak Helstrom family'') for $\{\omega^{M}_{i}\}_{i=1}^M$ such that 
\begin{align}
	\label{eq:Helstrom1}
	\tilde{p}_{i}=\tilde{p}_{j}\ge\frac{1}{M}
\end{align}
and
\begin{align}
	\label{eq:Helstrom2}
	\tilde{p}_{i}\omega^{M}_{i}+(1-\tilde{p}_{i})t_{i}=\tilde{p}_{j}\omega^{M}_{j}+(1-\tilde{p}_{j})t_{j},
\end{align}
for all $i,j=1, \ldots, M$, 
then $P_{M}\le\frac{1}{M\tilde{p}_{i}}$ holds. 
They also revealed that the equality holds if the observable $\{A_{i}\}_{i=1}^M$ satisfies $A_{i}(t_{i})=0$ for all $i$.
In the regular polygon theory $\Omega^{poly}_{M}$, we can easily find a weak Helstrom family for $\{\omega^{M}_{i}\}_{i=1}^M$.
In fact, if we set 
\begin{align}
	(t_{i}, \tilde{p}_{i})=
	\left\{
	\begin{aligned}
		&\left(\omega^{M}_{i+\frac{M}{2}},\ \frac{1}{2}\right)\quad\mbox{($M$: even)}\\
		&\left(\frac{\omega^{M}_{i+\frac{M-1}{2}}+\omega^{M}_{i+\frac{M+1}{2}}}{2},\  \frac{r_{M}^{2}}{1+r_{M}^{2}}\right)\quad\mbox{($M$: odd)},
	\end{aligned}
	\right.
\end{align}
then it is easy to see that they satisfy \eqref{eq:Helstrom1} and \eqref{eq:Helstrom2}.
Moreover, we can find that the condition $A_{i}(t_{i})=0$ is satisfied with each $A_{i}$ given by (remember \eqref{eq:polygon effect1} and \eqref{eq:polygon effect2})
\begin{align}
	A_{i}=
	\left\{
	\begin{aligned}
		&\frac{2}{M}e^{M}_{i}\quad\mbox{($M$: even)}\\
		&\frac{1+r_{M}^{2}}{M}e^{M}_{i}\quad\mbox{($M$: odd)}.
	\end{aligned}
	\right.
\end{align}
We note that the coefficients $\frac{2}{M}$ and $\frac{1+r_{M}^{2}}{M}$ are determined so that $\sum_{i=1}^MA_{i}=u_{sys}$ holds. 
This $\{A_{i}\}_{i}$ is realized by a channel $\Theta\in\mathcal{C}(\Omega_{tot})$ satisfying
\begin{align*}
	\Theta_{\pi_{i}(n)}^n=\Theta_{\pi_{i}(n)}^{n*}
	=
	\left\{
	\begin{aligned}
		&\frac{4}{M}\ket{e^{M}_{i}}\bra{e^{M}_{i}}\quad\mbox{($M$: even)}\\
		&\frac{(1+r_{M}^{2})^{2}}{M}\ket{e^{M}_{i}}\bra{e^{M}_{i}}\quad\mbox{($M$: odd)},
	\end{aligned}
	\right.
\end{align*}
where the linear operator $\ket{e^{M}_{i}}\bra{e^{M}_{i}}$ on $V=\real^3$ is defined as $\ket{e^{M}_{i}}\bra{e^{M}_{i}}\colon x\mapsto \braket{e^{M}_{i}|x}e^{M}_{i}$ with $\braket{\cdot|\cdot}$ the Euclidean inner product on $V$.
For this $\{A_{i}\}_{i}$, the probability $P_{suc}(\Theta)$ \eqref{eq:max prob0} attains its maximum value $P_M$ as
\begin{align}
	\label{eq:optimal for polygon}
	P_{M}=
	\left\{
	\begin{aligned}
		&\frac{2}{M}\quad\mbox{($M$: even)}\\
		&\frac{1+r_{M}^{2}}{M}\quad\mbox{($M$: odd)}.
	\end{aligned}
	\right.
\end{align}
We can see, for example, that $P_{3}=1$, 
and $P_{4}=\frac{1}{2}$.
The above result \eqref{eq:optimal for polygon} is compared with the success probability to program $M$ reversible dynamics on the system $\Delta_N$ by a single bit $\Delta_2$ (thus the total system is $\Delta_N \otimes \Delta_2$).
Since $\Delta_2$ has two pure states, it can program perfectly at most two reversible dynamics on $\Delta_N$.
It follows that in this case the optimal value of average success probability is $\frac{2}{M}$. 
This manifests that when implementing $M$ reversible dynamics on a classical system, using the regular polygon theory as apparatus with $M$ sides results in a better or equal success probability than using classical bit.

\section{Conclusion}
In this study, we considered the generalization of quantum programming scheme to GPTs.
It was found that a family of reversible dynamics 
on a target system is programmable only if a pairwise distinguishable set of states in an apparatus is used as programs.
While this result seems to be just a straightforward generalization of the quantum result, it should be emphasized that this was obtained for any physically valid composite of the system and apparatus, i.e., any bipartite state space between the minimal and the maximal tensor products of them.
On the other hand, we also considered changing the programming scenario itself: universal programming of reversible dynamics and channels, and investigated when an apparatus makes them possible.
It was demonstrated that the former scheme is realizable if and only if the apparatus has a quasi-classical structure, which was originally derived for the target system in the initial programming scheme, and that the latter is possible if and only if the corresponding programs in the apparatus are perfectly distinguishable.
We believe that the former result is particularly important in that it is peculiar to GPTs beyond quantum theory.
We also presented numerical evaluations for how well we can implement channels on a classical system if states in regular polygon theories (that are in general not perfectly distinguishable) are used as programs, where only approximate programming is possible.
It will be interesting to present similar evaluations for more general cases when channels on non-classical GPTs are to be programmed.
Future study will be also needed to give further investigations of quasi-classical structures.
As Example \ref{eg:triangular prism} shows, state spaces with quasi-classical substructures seem to have properties that the classical (triangle) and square theories have.
Since these two theories exhibit respectively minimum and maximum values for the CHSH value \cite{PhysRevA.75.032304,PR-box} or incompatibility \cite{PhysRevA.96.022113}, it may be possible to give other theoretical characterizations for quasi-classical structures.

\section*{Acknowledgment} 
The authors thank Yui Kuramochi for helpful suggestions.
TM acknowledges financial support from JSPS (KAKENHI Grant No. JP20K03732).
RT acknowledges financial support from JSPS (KAKENHI Grant No. JP21J10096).

\appendix
\def\thesection{Appendix\ \Alph{section}}
\section{A pairwise distinguishable family of states is always finite.}
\label{appA}
\renewcommand{\theequation}{A.\arabic{equation}}
\setcounter{equation}{0}
\renewcommand{\thesection}{\Alph{section}}
\setcounter{subsection}{0}
In this appendix, we prove that a pairwise distinguishable family of states in a finite-dimensional state space is a finite set.
\begin{prop}
	\label{prop:appA}
	Let $\Omega$ be a state space in $V=\mathit{span}(\Omega)$ with $\dim V=d$ $(d<\infty)$, and let $C:=\{\omega_i\}_{i\in\mathcal{I}}\subset\Omega$ be a pairwise distinguishable set of states.
	Then $C$ is a finite set, i.e., $|\mathcal{I}|<\infty$.
\end{prop}
\begin{pf}
	Since $\dim\mathit{span}(\Omega)=d$, we can choose a linearly independent set of states $\{\tilde{\omega}_k\}_{k=1}^d$ that forms a basis of $V$.
	Then a set of linear functions $\{w_{k}\}_{k=1}^d$ on $V$ defined as $w_k(\tilde{\omega}_{k'})=\delta_{kk'}$ can be introduced.
	It can be seen that $\{w_{k}\}_{k=1}^d$ is a linearly independent set of vectors, and thus forms a basis of the dual space $V^*$.
	For such vectors, we define $s:=\min_{k\in\{1, \ldots, d\}}\inf_{\omega\in\Omega} w_k(\omega)$ and $t:=\max_{k\in\{1, \ldots, d\}}\sup_{\omega\in\Omega} w_k(\omega)$.
	We note that $s$ and $t$ are finite quantities because $\Omega$ is a compact set.
	Now we divide the interval $[s, t]$ into 
	$M$ parts as $A_1:= [s,s+\frac{t-s}{M}),\ A_2:=[s+\frac{t-s}{M}, s+2\frac{t-s}{M}),\ \cdots,\ A_M:=[t-\frac{t-s}{M}, t]$.
	It defines a pairwise disjoint partition of $\Omega$ by 
	\[
	\Omega=\bigcup_{(n_1, n_2, \ldots, n_d)\in \{1,2,\ldots, M\}^d}P_{n_1 n_2 \ldots n_d},
	\]
	where $P_{n_1 n_2 \ldots n_d}:= \Omega\cap w_{1}^{-1}(A_{w_1}) \cap 
	w_{2}^{-1} (A_{n_2}) \cap \cdots\cap w_{d}^{-1}(A_{n_d})$.
	In other words, a state $\sigma \in P_{n_1, n_2, \ldots, n_d}$ satisfies $w_{k}(\sigma) \in A_{n_k}$ for each $k=1, \ldots, d$. 
	Let us suppose that the set $C$ is an infinite set.
	Because $C$ is infinite, there exists at least one $P_{n_1 n_2 \ldots n_d}$ such that $P_{n_1 n_2 \ldots n_d}\cap C$ is an infinite set ($|P_{n_1 n_2  \ldots  n_d}\cap C|\geq 2$ is enough for the following argument).  
	For such $P_{n_1 n_2 \ldots n_d}\cap C$, we take its elements $\sigma_1, \sigma_2$.
	Since $\{\sigma_1, \sigma_2\}$ are perfectly distinguishable, 
	there exists an effect $e$ such that $e(\sigma_1)=1$ and $e(\sigma_2)=0$ holds. 
	On the other hand, representing $e\in V^*$ in terms of the basis $\{w_k\}_k$ as $e=\sum_{k=1}^d\alpha_k w_k$ $(\alpha_k\in\real)$, we have $\alpha_k\in [0, 1]$ for each $k=1, \ldots, d$ because $e(\tilde{\omega}_k)\in [0, 1]$ holds.
It follows that 
	\begin{align*}
		1=|e(\sigma_1 - \sigma_2)| 
		=\left|\sum_{k=1}^d \alpha_ k w_k(\sigma_1 - \sigma_2)\right|
		\leq \frac{t-s}{M}\sum_{k=1}^d |\alpha_k |
		\le\frac{d(t-s)}{M}, 
	\end{align*}
	but, because we can choose arbitrarily large $M$, 
	it contradicts.\qed
\end{pf}
\begin{rmk}
For a system described by a state space $\Omega$, the integer
\[
d_i:=\max\{t\in\mathbb{N}\mid\mbox{$\{\omega_1, \ldots, \omega_{t}\}\subset\Omega$ is pairwise distinguishable}\}
\] 
is called the \textit{information dimension} of the system \cite{Brunner_2014}.
Proposition \ref{prop:appA} indicates that the information dimension of a system with a finite-dimensional state space is always finite (thus $d_i$ is a well-defined quantity).
\end{rmk}

\appendix
\setcounter{section}{1}
\def\thesection{Appendix\ \Alph{section}}
\section{$N$ distinct states can be discriminated by an $N$-outcome observable via the observed statistics}
\label{appB}
\renewcommand{\theequation}{B.\arabic{equation}}
\setcounter{equation}{0}
\renewcommand{\thesection}{\Alph{section}}
\setcounter{subsection}{0}
In this appendix, we prove that $N$ distinct states can be discriminated by an observable with $N$ outcomes in terms of the observed probability distributions. 
\begin{prop}
	\label{prop:appB}
Let $\Omega$ be a state space in $V=\mathit{span}(\Omega)$ with $\dim V=d\ (<\infty)$, and let $D:=\{\omega_n\}_{n=1}^N$ $(N<\infty)$ be a finite set of distinct states.
There exists an observable $F=\{f_n\}_{n=1}^N$ with $N$ outcomes such that the observed probability distributions $p(\cdot|\omega_n):=\{\ang{f_{n'}, \omega_n}\}_{n'=1}^N$ $(n=1, \ldots, N)$ are all distinct.
\end{prop}
\begin{pf}
Let $V':=\mathit{span}(D)$ be a subspace of $V$, and let $M:=\dim V'\ (M\le N)$ be its dimension.
Then there exists a linearly independent set  $\{\tilde{\omega}_{m}\}_{m=1}^M\subseteq D$.
Also, there is a linearly independent set of states $\{\tilde{\omega}_{l}\}_{l=m+1}^d\subseteq\Omega\backslash D$ such that $\{\tilde{\omega}_{k}\}_{k=1}^d$ forms a basis of $V$.
We introduce a set $\{w_k\}_{k=1}^d$ of elements of $V^*$ defined as $w_k(\tilde{\omega}_{k'})=\delta_{kk'}$, which is a basis of $V^*$.
It can be seen easily that the unit effect $u\in V^*$ for $\Omega$ is represented as $u=\sum_{k=1}^dw_k$.
Now we construct an $M$-outcome observable that discriminates states in $V'$.
Let $\{b_m\}_{m=1}^M$ be a set of elements in $V^*$ defined as
\[
b_m=
\left\{
\begin{aligned}
&\ \ w_m&&\quad(m=1, \ldots, M-1)\\
&\sum_{l=M}^dw_l&&\quad(m=M).
\end{aligned}
\right.
\]
We note that $\sum_{m=1}^Mb_m=\sum_{k=1}^dw_k=u$ holds.
The vectors $\{b_m\}_{m=1}^M$ discriminates states in $V'\cap \Omega$.
In fact, if states $\omega,\omega'\in V'\cap \Omega$ satisfy $b_m(\omega)=b_m(\omega')$ for all $m=1, \ldots, M$, then, representing the states as $\omega=\sum_{m=1}^M\alpha_m\tilde{\omega}_m$ and $\omega'=\sum_{m=1}^M\beta_m\tilde{\omega}_m$ via the basis $\{\tilde{\omega}_{m}\}_{m=1}^M$ of $V'$, we obtain $\alpha_m=\beta_m$ for all $m=1, \ldots, M$, that is, $\omega=\omega'$.
Let $c=\inf_{m\in\{1, \ldots, M\}}\inf_{\omega\in\Omega}b_m(\omega)$.
We have $[b_m-cu](\omega)\ge0$ for all $\omega\in\Omega$ and $m=1, \ldots, M$.
In addition, the vectors $\{b_m-cu\}_{m=1}^M$ discriminate states in $V'\cap \Omega$ because the condition $b_m(\omega)=b_m(\omega')$ is equivalent to $[b_m-cu](\omega)=[b_m-cu](\omega')$ whenever $\omega, \omega'\in\Omega$.
Then we take a suitable normalization for $\{b_m-cu\}_{m=1}^M$ to obtain an observable $E=\{e_m\}_{m=1}^M$ that discriminates states in $D$ (remember that $\sum_{m=1}^Mb_m=u$ holds). 
Adding the zero effect to $E$ if necessary, we have an $N$-outcome observable $F$ that discriminates states in $D$. \qed
\end{pf}
\begin{rmk}
From the above proof, we obtain the following observation:
for a state space $\Omega$, there exist observables with at most $\dim \textit{span}(\Omega)$ outcomes such that an arbitrary number of distinct states can be discriminated through the probability distributions observed in their measurements.
Those observables are called \textit{informationally complete observables} \cite{Busch_informationally_complete,doi:10.1063/1.529975}, and informationally complete observables with $\dim \textit{span}(\Omega)$ outcomes (i.e. observables whose effects compose a basis of $\textit{span}(\Omega)$) are called particularly \textit{minimal informationally complete observables} \cite{Barnum_nocloning}. 
\end{rmk}

\appendix
\setcounter{section}{2}
\def\thesection{Appendix\ \Alph{section}}
\section{Proof of Proposition \ref{classicalp}}
\label{appC}
\renewcommand{\theequation}{C.\arabic{equation}}
\setcounter{equation}{0}
\renewcommand{\thesection}{\Alph{section}}
\setcounter{subsection}{0}
In this appendix, we present the proof of Proposition \ref{classicalp} shown as follows.
\begin{propapp}
	Let $\Omega$ and $\Omega^{ext}$ be a state space and the set of all pure states respectively, and let $\Omega^{ext}=\bigcup_{z\in \mathcal{Z}}
	\Omega^{ext}[z]$ be a disjoint decomposition for $\Omega^{ext}$. 
	The decomposition $\Omega^{ext}=\bigcup_{z\in \mathcal{Z}}
	\Omega^{ext}[z]$ is quasi-classical if and only if it satisfies the following condition ($\star$):\\
($\star$)
	if $\omega \in \Omega$ is expressed as $\omega= \sum_{z\in \mathcal{Z}}
	p_z \omega_z=\sum_{z\in \mathcal{Z}}
	q_z \omega'_z$, where each $\omega_z, \omega'_z \in 
	\Omega[z]$ with $\Omega[z]$ the convex hull of $\Omega^{ext}[z]$ and $\{p_z\}_{z\in\mathcal{Z}}$ and $\{q_z\}_{z\in\mathcal{Z}}$ probability 
	distributions on $\mathcal{Z}$, then $p_z=q_z$ holds for all $z\in Z$.
\end{propapp}
\begin{pf}
(\textit{the `only if' part})\\
	Due to the Krein-Milman theorem \cite{Conway_functionalanalysis}, any $\omega\in \Omega$ can be 
	decomposed as 
	$\omega=\sum_{z}\sum_{j} 
	\lambda_{z, j}\omega_{z, j}$, where 
	$\omega_{z,j}\in \Omega^{ext}[z]$ 
	and $\{\lambda_{z,j}\}_{z,j}$ is a probability distribution.
	We introduce $\omega_z=\frac{\sum_j \lambda_{z,j}\omega_{z,j}}{\sum_j \lambda_{z,j}}\in \Omega[z]$ 
	and
	$p_z= \sum_j \lambda_{z,j}$ for $z\in \mathcal{Z}$ satisfying $\sum_j \lambda_{z,j} \neq 0$ to obtain 
	$\omega= \sum_z p_z \omega_z$. 
	To prove the uniqueness of $\{p_z\}_z$, suppose that there are two decomposition 
	$\omega= \sum_z p_z \omega_z 
	= \sum_z q_z \sigma_z$ with $\omega_z, \sigma_z
	\in \Omega[z]$. 
	Applying the quasi-classical observable $\{A_z\}_z$ associated with the quasi-classical decomposition, we obtain 
	$p_z = q_z$ for each $z\in\mathcal{Z}$. \\
	(\textit{the `if' part})\\
The claim is proved by induction on the integer $|\mathcal{Z}|$.
First we consider the case of $|\mathcal{Z}|=2$, i.e., $\mathcal{Z}=\{z_1, z_2\}$.
For the corresponding decomposition $\Omega^{ext}=\Omega^{ext}[z_1]\cup\Omega^{ext}[z_1]$, it holds that  $\textit{aff}(\Omega^{ext}[z_1])\cap\textit{aff}(\Omega^{ext}[z_2])=\emptyset$.
To see this, suppose that $\textit{aff}(\Omega^{ext}[z_1])\cap\textit{aff}(\Omega^{ext}[z_2])\neq\emptyset$ and $v\in\textit{aff}(\Omega^{ext}[z_1])\cap\textit{aff}(\Omega^{ext}[z_2])$.
There exist positive numbers $\{c_i^+\}_{i}$, $\{c_j^-\}_{j}$, $\{d_k^+\}_{k}$, and $\{d_l^-\}_l$, and pure states $\{\omega^{1+}_i\}_{i}\subseteq\Omega^{ext}[z_1]$, $\{\omega^{1-}_j\}_{j}\subseteq\Omega^{ext}[z_1]$, $\{\omega^{2+}_k\}_{k}\subseteq\Omega^{ext}[z_2]$, and $\{\omega^{2-}_l\}_{l}\subseteq\Omega^{ext}[z_2]$ such that 
\begin{align}
&\sum_ic^+_i-\sum_jc^-_j=1,\quad \sum_kd^+_k-\sum_ld^-_l=1,\label{eq:appC1}\\
&v=\sum_ic^+_i\omega^{1+}_i-\sum_jc_j^-\omega^{1-}_j
=\sum_kd^+_k\omega^{2+}_k-\sum_ld_l^-\omega^{2-}_l.
\end{align}
It follows that 
\[
\sum_ic^+_i\omega^{1+}_i+\sum_ld_l^-\omega^{2-}_l
=\sum_jc_j^-\omega^{1-}_j+\sum_kd^+_k\omega^{2+}_k,
\]
or
\begin{align*}
\frac{\sum_ic^+_i}{K}\cdot\frac{\sum_ic^+_i\omega^{1+}_i}{\sum_ic^+_i}
&+
\frac{\sum_ld_l^-}{K}\cdot\frac{\sum_ld_l^-\omega^{2-}_l}{\sum_ld_l^-}\\
&\ \ =\frac{\sum_jc^-_j}{K}\cdot\frac{\sum_jc^-_j\omega^{1-}_j}{\sum_jc^-_j}
+
\frac{\sum_kd^+_k}{K}\cdot\frac{\sum_kd^+_k\omega^{2+}_k}{\sum_kd^+_k},
\end{align*}
where $K=\sum_ic^+_i+\sum_ld_l^-=\sum_jc^-_j+\sum_kd^+_k(>0)$.
Because 
\[
\frac{\sum_ic^+_i\omega^{1+}_i}{\sum_ic^+_i},\ \frac{\sum_jc^-_j\omega^{1-}_j}{\sum_jc^-_j}
\in\Omega[z_1]
\quad\mbox{and}\quad
\frac{\sum_kd^+_k\omega^{2+}_k}{\sum_kd^+_k},\ 
\frac{\sum_ld_l^-\omega^{2-}_l}{\sum_ld_l^-}
\in\Omega[z_2]
\] 
hold, we have from the assumption $(\star)$
\[
\sum_ic^+_i=\sum_jc^-_j,\ \ \sum_kd^+_k=\sum_ld_l^-.
\]
This contradicts \eqref{eq:appC1}, and thus $\textit{aff}(\Omega^{ext}[z_1])\cap\textit{aff}(\Omega^{ext}[z_2])=\emptyset$ is concluded.
Then, according to the separating hyperplane theorem (Theorem 11.2 in \cite{Rockafellar+2015}), there exist hyperplanes $H_1$ and $H_2$ in $V:=\textit{span}(\Omega)$ such that
$\textit{aff}(\Omega^{ext}[z_1])\subseteq H_1$, $\textit{aff}(\Omega^{ext}[z_2])\subseteq H_2$, and $H_1\cap H_2=\emptyset$ (in particular, $H_1$ and $H_2$ are parallel).
They can be represented as $H_1$ and $H_2$ as $H_1=\{x\in V\mid \ang{x, h}=a\}$ and $H_1=\{x\in V\mid \ang{x, h}=b\}$ respectively, where $a, b\in\real$ with $a<b$ and $h\in V$, and $\ang{\cdot, \cdot}$ is an inner product in $V$. 
We note that we identify the dual space $V^*$ with $V$ via the inner product $\ang{\cdot, \cdot}$ in the following.
We can observe that the vector $e:=\frac{1}{b-a}(h-au)$, where $u$ is the unit effect, satisfies 
\begin{align}
	\label{eq:appC2}
\ang{e, \omega^{ext}}=
\left\{
\begin{aligned}
&0\quad(\omega^{ext}\in\Omega^{ext}[z_1])\\
&1\quad(\omega^{ext}\in\Omega^{ext}[z_2]).
\end{aligned}
\right.
\end{align}
In particular, $e$ is properly an effect because $\ang{e, \omega}\in[0, 1]$ for all $\omega\in\Omega$ due to \eqref{eq:appC2}, and thus $\Omega^{ext}=\Omega^{ext}[z_1]\cup\Omega^{ext}[z_1]$ is quasi-classical ($\{e, u-e\}$ gives the corresponding quasi-classical observable).

Assume that the claim holds when $|\mathcal{Z}|=N\ (N\ge2)$, and consider the decomposition $\Omega^{ext}=\bigcup_{z\in \mathcal{Z}}
\Omega^{ext}[z]$ with $|\mathcal{Z}|=N+1$ satisfying the relevant condition $(\star)$.
Expressing $\mathcal{Z}=\{z_1, \ldots, z_{N+1}\}$, we introduce the disjoint decomposition $\Omega^{ext}=\bigcup_{i=1}^{N}
\Omega_{1}^{ext}[z_i]$, where
\[
\Omega_{1}^{ext}[z_i]=
\left\{
\begin{aligned}
&\Omega^{ext}[z_i]\quad&&(i=1, \ldots, N-1)\\
&\Omega^{ext}[z_{N}]\cup\Omega^{ext}[z_{N+1}]\quad&&(i=N).
\end{aligned}	
\right.
\]
It is not difficult to see that this decomposition satisfies $(\star)$, and thus there exists a quasi-classical observable $A=\{A_i\}_{i=1}^N$ such that 
\begin{align*}
A_i(\omega)=
\left\{
\begin{aligned}
	&1\quad(\omega\in\Omega_{1}^{ext}[z_i])\\
	&0\quad(\omega\in\Omega_{1}^{ext}[z_j]\ \mbox{with}\ j\neq i).
\end{aligned}	
\right.
\end{align*}
On the other hand, if we consider the disjoint decomposition 
$\Omega^{ext}=\bigcup_{i=1}^{N}
\Omega_{2}^{ext}[z_i]$, where
\[
\Omega_{2}^{ext}[z_i]=
\left\{
\begin{aligned}
	&\Omega^{ext}[z_{i+1}]\quad&&(i=1, \ldots, N-1)\\
	&\Omega^{ext}[z_{N+1}]\cup\Omega^{ext}[z_{1}]\quad&&(i=N),
\end{aligned}	
\right.
\]
then there similarly exists an observable $B=\{B_i\}_{i=1}^N$ such that 
\begin{align*}
	B_i(\omega)=
	\left\{
	\begin{aligned}
		&1\quad(\omega\in\Omega_{2}^{ext}[z_i])\\
		&0\quad(\omega\in\Omega_{2}^{ext}[z_j]\ \mbox{with}\ j\neq i).
	\end{aligned}	
	\right.
\end{align*}
Now we obtain easily $A_{i+1}=B_{i}$ for $i=1, \ldots, N-1$.
Besides, we can observe that $(B_{N}-A_{1})$ is a proper effect satisfying 
\begin{align*}
	(B_{N}-A_{1})(\omega)=
	\left\{
	\begin{aligned}
		&1\quad(\omega\in\Omega^{ext}[z_{N+1}])\\
		&0\quad(\omega\in\Omega_{2}^{ext}[z_i]\ \mbox{with}\ i\neq N+1).
	\end{aligned}	
	\right.
\end{align*}
Therefore, a family $C=\{C\}_{i=1}^{N+1}$ of effects defined as
\begin{align*}
	C_i=
	\left\{
	\begin{aligned}
		&A_i&&\quad(i=1, \ldots, N-1)\\
		&B_{N-1}&&\quad(i=N)\\
		&B_N-A_1&&\quad(i=N+1).
	\end{aligned}	
	\right.
\end{align*}
is an observable that satisfies 
\begin{align*}
	C_i(\omega)=
	\left\{
	\begin{aligned}
		&1\quad(\omega\in\Omega^{ext}[z_i])\\
		&0\quad(\omega\in\Omega^{ext}[z_j]\ \mbox{with}\ j\neq i),
	\end{aligned}	
	\right.
\end{align*}
that is, the initially considered decomposition $\Omega^{ext}=\bigcup_{z\in \mathcal{Z}}
\Omega^{ext}[z]$ is quasi-classical.\qed

\end{pf}

\bibliographystyle{hieeetr_url} 
\bibliography{ref_program_0310}

\end{document}